\documentclass[aps,pra,reprint,showpacks,twocolumn,superscriptaddress]{revtex4-1}%
\usepackage[colorlinks,linkcolor=blue,citecolor=blue,urlcolor=blue,hyperindex,bookmarks=false,pdfstartview=FitH]{hyperref}
\usepackage{amsfonts}
\usepackage{mathrsfs}
\usepackage{amsmath}
\usepackage{amssymb}
\usepackage{graphicx}
\usepackage{dcolumn}
\usepackage{color}%
\usepackage{bm}
\usepackage{soul}
\usepackage[left]{lineno}
\usepackage{ulem}
\usepackage{changepage}
\providecommand{\U}[1]{\protect\rule{.1in}{.1in}}

\newcommand{\figpanel}[2]{\hyperref[#1]{\ref*{#1}(#2)}}
\begin{document}

\title{Single-photon manipulations based on optically-controlled chiral couplings in waveguide structures of Rydberg giant atoms}
\author{Yao-Tong Chen}
\affiliation{School of Physics and Center for Quantum Sciences, Northeast Normal University, Changchun 130024, China}
\author{Lei Du}
\affiliation{Department of Microtechnology and Nanoscience, Chalmers University of Technology, 41296 Gothenburg, Sweden}
\author{Zhihai Wang}
\affiliation{School of Physics and Center for Quantum Sciences, Northeast Normal University, Changchun 130024, China}
\author{M. Artoni}
\email{maurizio.artoni@unibs.it}
\affiliation{Department of Engineering and Information Technology, Brescia University, 25133 Brescia, Italy}
\affiliation{European Laboratory for Nonlinear Spectroscopy (LENS), 50019 Sesto Fiorentino, Italy}
\author{G. C. La Rocca}
\email{giuseppe.larocca@sns.it}
\affiliation{NEST, Scuola Normale Superiore, 56126 Pisa, Italy}
\author{Jin-Hui Wu}
\email{jhwu@nenu.edu.cn}
\affiliation{School of Physics and Center for Quantum Sciences, Northeast Normal University, Changchun 130024, China}
\date{\today}

\begin{abstract}
Two interacting Rydberg atoms coupled to a waveguide realize a giant-atom platform that exhibits the controllable (phase-dependent) chirality where the direction of nonreciprocal photon scattering can be switched on demand, e.g., by the geometrical tuning of an external driving field. At variance with previous chiral setups, the simplified approach of our proposed platform arises from an optical implementation of the local phase difference between two coupling points of the Rydberg giant atom. Furthermore, employing two or more driving fields, this platform could also be used as a frequency converter with its efficiency exhibiting a strong asymmetry and being significantly enhanced via the chiral couplings. Our results suggest an extendable giant-atom platform that is both innovative and promising for chiral quantum optics and tunable frequency conversion in the optical domain.
\end{abstract}

\maketitle
\section{Introduction}

Rydberg atoms with large principal quantum numbers, combining long coherence times and strong long-range interactions, can serve as attractive building blocks for many important applications, including scalable quantum computing and long-distance quantum communication~\cite{Rydreview,Ryd1,Ryd2}. Note, in particular, that strong Rydberg interactions can well suppress multiple atomic excitations within a blockade radius by shifting the resonance of double atomic excitations~\cite{blockade}, which opens the possibilities to explore single-photon generation~\cite{source1,source2,source3}, quantum logic gates \cite{gate1,gate2,gate3}, entangled states~\cite{en1,en2,en3,en4}, and quantum simulators~\cite{sim1,sim2}. Very recently, interacting Rydberg atoms have been considered as a new platform to implement actual giant-atom physics working in the optical domain, with peculiar self-interference effects and entanglement-onset dynamics~\cite{RydGiant}. It is then worth noting that giant atoms have emerged as a novel paradigm in quantum optics and are generally characterized by multiple couplings with electromagnetic or acoustic modes at distinct points, hence breaking the dipole approximation~\cite{GiantReview}. Experimental platforms of typical giant atoms are presently available including, e.g., superconducting quantum circuits~\cite{saw14,2020nature,decay2,Wilson2021}, coupled waveguide arrays~\cite{longhiretard}, and ferromagnetic spin systems~\cite{fer}.

Benefiting from diverse geometric structures owing to multiple coupling points between giant atoms and waveguide modes, rich interference effects can be introduced to modify relevant interactions, opening a broad field of perspectives for controlling photon transport and information processing.
A series of unique characteristics, e.g., frequency-dependent atomic relaxation rates and Lamb shifts~\cite{lxy,2014Kockum,continuum1}, non-exponential atomic decay~\cite{decay1,decay2}, in-band decoherence-free interactions~\cite{2020nature,free,fra,AFKchiral}, and long-lived entanglement generation~\cite{2023en}, have already been discovered in different giant-atom schemes.
Especially, for some chiral setups based on giant atoms, unlike others based on spin-momentum locking effect~\cite{spin1,spin2,spin3} or topological waveguides~\cite{topological1,topological2}, the direction-dependent couplings can be realized by introducing a local coupling-phase difference to break the time-reversal symmetry, resulting in chiral spontaneous emission and nonreciprocal transmission~\cite{WXarxiv,DLprl,cp2022,prx2023}.
Considering specifically designed multi-level atomic structures in waveguide quantum electrodynamics (QED) systems, more abundant behaviors of photon scattering will appear, including asymmetric photon routing~\cite{router2,router3} or circulating~\cite{circulator1,circulator2}, and efficient frequency conversion~\cite{sag,jia2017,bethe2,Wang2014}.

\begin{figure}[pth]
\centering
\includegraphics[width=8.5 cm]{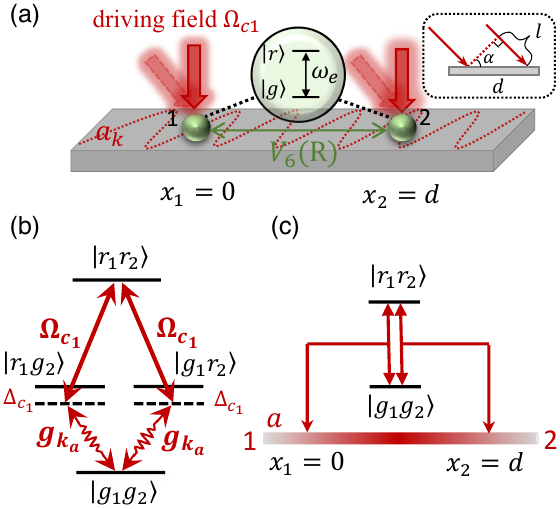}
\caption{Schematic diagram for achieving nonreciprocal transmission of one waveguide mode. (a) Two two-level Rydberg atoms interacting via a vdW potential $V_{6}$, coupled to a waveguide mode $a$ respectively at $x_{1}=0$ and $x_{2}=d$, and driven by a coherent field $\Omega_{c1}$. If the driving field is oblique with angle $\alpha$, there will be a phase difference as it reaches the two atoms ascribed to optical path difference $l$ (see the inset). (b) The four-level configuration in the two-atom basis where driving field $\Omega_{c1}$ is matched in frequency with waveguide mode $a$ on both left and right paths of two-photon resonance. (c) The equivalent two-level giant atom exhibiting two coupling points when adiabatically eliminating two single-excitation states.}\label{fig1}
\end{figure}

This paper aims at tackling a few issues of the optical giant-atom physics using Rydberg atoms, building upon our earlier work~\cite{RydGiant}. We extend this work to encompass scenarios involving multiple drivings and chiral couplings to attain on demand nonreciprocal light scattering and asymmetric frequency conversion. In Sect.~II, we consider the basic case of two two-level Rydberg atoms coupled to one waveguide mode and driven by one external field (see Fig.~\ref{fig1}). In an appropriate regime, both single-excitation states remain unpopulated and can be adiabatically eliminated, leading to an equivalent giant atom coupled to the waveguide mode at two points. Scattering properties of this giant atom are sensitive to the relative phase of two waveguide-atom coupling coefficients and also controlled by the driving field's angle of incidence, thus yielding tunable nonreciprocal transmissivities.

In Sect.~III, we consider another case in which different Rydberg atoms are driven by different external fields matching different waveguide modes in frequency under the two-photon resonance (see Fig.~\ref{fig4}). While this nonlocal system with each waveguide mode exhibiting one coupling point no longer mimics a giant atom, hence exhibits reciprocal scattering properties, it may operate as a symmetric frequency converter whereby a photon propagating in one waveguide mode, independent of its input port, can be converted into a photon propagating in the other waveguide mode and vice versa. In Sect. IV, we consider the more complex case in which both Rydberg atoms are driven by two external fields and coupled to two waveguide modes with matched frequencies under the two-photon resonance (see Fig.~\ref{fig6}). This nonlocal system with each waveguide mode exhibiting two coupling points behaves as a giant atom again and becomes sensitive to two relative phases of four waveguide-atom coupling coefficients. Its frequency conversion properties can be made highly asymmetric and more efficient, e.g., by changing the angle of incidence of each driving field.

\section{nonreciprocal transmission}

We consider two identical atoms $1$ and $2$ with resonant transition frequency $\omega_{e}$ and intrinsic dissipation rate $\gamma$ (into the free space) from Rydberg state $|r\rangle$ to ground state $|g\rangle$. They are driven by a coherent field with Rabi frequency $\Omega_{c1}$ and coupled to a photonic crystal waveguide at $x_{1}=0$ and $x_{2}=d$, respectively, as shown in Fig.~\figpanel{fig1}{a}. They also interact via a van der Waals (vdW) potential $V_{6}=C_{6}/R^{6}$~\cite{Rydreview}, where $C_{6}$ and $R$ are the vdW coefficient and the interatomic distance, respectively.
This potential shifts the frequency of double-excitation state $|r_{1}r_{2}\rangle$ from $2\omega_{e}$ to $2\omega_{e}+V_{6}$ yet without affecting the frequency $\omega_{e}$ of single-excitation states $|r_{1}g_{2}\rangle$ and $|g_{1}r_{2}\rangle$. The driving field is assumed to exhibit a frequency $\omega_{c1}$ close to $\omega_{e}+V_{6}$ but sufficiently far detuned from $\omega_{e}$. We further assume that $\omega_{e}+V_{6}$ falls within a band gap of the waveguide, while $\omega_{e}$ corresponds to a propagating waveguide mode $a$. Then, the two upper transitions  $|r_{1}g_{2}\rangle\rightarrow|r_{1}r_{2}\rangle$ and $|g_{1}r_{2}\rangle\rightarrow|r_{1}r_{2}\rangle$ only couple with the driving field $\Omega_{c1}$, while the two lower transitions $|g_{1}g_{2}\rangle\rightarrow|r_{1}g_{2}\rangle$ and $|g_{1}g_{2}\rangle\rightarrow|g_{1}r_{2}\rangle$ only couple with the waveguide mode $a$. Based on all above considerations, as done in our recent work~\cite{RydGiant}, we can obtain a four-level configuration in the two-atom basis as shown in Fig.~\figpanel{fig1}{b} where the detuning $\Delta_{c1}=\omega_{c1}-(\omega_{e}+V_{6})$ is restricted by $|\Delta_{c1}|\ll V_6$ and the resonance condition on two-photon transition $|g_{1}g_{2}\rangle\rightarrow|r_{1}r_{2}\rangle$ can be satisfied for a matching waveguide-mode frequency $\omega_{e}-\Delta_{c1}$.

The system Hamiltonian, under the rotating-wave approximation, can be written as ($\hbar=1$)
\begin{equation}
\begin{split}
H_{A_{k}}&=(\omega_{e}-i\gamma)(|g_{1}r_{2}\rangle\langle g_{1}r_{2}|+|r_{1}g_{2}\rangle\langle r_{1}g_{2}|)\\
&+(2\omega_{e}+V_{6}-2i\gamma)|r_{1}r_{2}\rangle\langle r_{1}r_{2}|
+\int dk_{a}\omega_{ka}a_{k}^{\dagger}a_{k}\\
&+\Big[\int g_{a}dk_{a}a_{k}(|r_{1}g_{2} \rangle\langle g_{1}g_{2}|+|g_{1}r_{2} \rangle\langle g_{1}g_{2}|e^{ik_{a}d})\\
&+\Omega_{c1} e^{-i\omega_{c1}t}(|r_{1}r_{2} \rangle\langle r_{1}g_{2}|+e^{i\theta_{1}}|r_{1}r_{2} \rangle\langle g_{1}r_{2}|)+\text{H.c.}\Big].
\label{HA_k}
\end{split}
\end{equation}
Here, $a_{k}$ ($a_{k}^{\dagger}$) refers to the bosonic annihilation (creation) operators of waveguide mode $a$ denoted by wave vector $k_{a}$ and frequency $\omega_{ka}$; $\theta_{1}$ represents the phase difference between two Rabi frequencies of a common magnitude $\Omega_{c1}$ for different atoms, which is easily controlled by utilizing an oblique driving field with angle $\alpha$ deviating from the normal to the waveguide. To be more concrete, we have $\theta_{1}=k_{c1}l_{1}$ where $k_{c1}$ is the wave vector of the driving field and $l_{1}=d\sin\alpha$ is the optical path difference between the two driving points (being $\theta_{1}=0$ when the field is incident normally).
Under the Weisskopf-Wigner approximation, we further have a constant coupling strength $g_{ka}=g_{a}$ for the waveguide modes of frequencies $\omega_{ka}\simeq\omega_{e}$ and a constant atomic decay rate $\Gamma_{a}=g_{a}^{2}/v_{g}$ into these waveguide modes with $v_{g}$ being the group velocity of waveguide photons. The present model differs from the model in~\cite{RydGiant} for the relative phase $\theta_{1}$, which would bring chirality into play as discussed later. Moreover, we investigate here the scattering of a single waveguide photon by the two Rydberg atoms, initially in the ground state $|g_{1}g_{2}\rangle$, while omitting spontaneous emission following the initial Rydberg state $|r_{1}r_{2}\rangle$ as done in~\cite{RydGiant}.

We will employ the approach developed by Shen and Fan~\cite{fan1,fan2}, used in~\cite{bethe1,bethe2,bethe3,cp2022}, and referred to as the Bethe-ansatz approach~\cite{liao2016}. Thus, the real-space description of our model is required with the corresponding Hamiltonian transferred from $H_{A_k}$ being
\begin{equation}
\begin{split}
H_{A_{x}}&=(\omega_{e}-i\gamma)(|r_{1}g_{2} \rangle\langle r_{1}g_{2}|+|g_{1}r_{2} \rangle\langle g_{1}r_{2}|)\\
&+(2\omega_{e}+V_{6}-2i\gamma)|r_{1}r_{2}\rangle\langle r_{1}r_{2}|\\
&+\int_{-\infty}^{+\infty}dx\Big\lbrack a_{L}^{\prime\dagger}(x)\Big(\omega_{0}+iv_{g}\frac{\partial} {\partial x}\Big)a_{L}^{\prime}(x)\\
&+a_{R}^{\prime\dagger}(x)\Big(\omega_{0}-iv_{g}\frac{\partial} {\partial x}\Big)a_{R}^{\prime}(x)\Big\rbrack\\
&+\int_{-\infty}^{+\infty}dx\Big\lbrace\delta(x)g_{a}[a_{R}(x)+a_{L}(x)]|r_{1}g_{2} \rangle\langle g_{1}g_{2}|\\
&+\delta(x-d)g_{a}\big\lbrack a_{R}(x)+a_{L}(x)\big\rbrack|g_{1}r_{2} \rangle\langle g_{1}g_{2}|+\text{H.c.}\Big\rbrace\\
&+\Omega_{c1}e^{-i\omega_{c1}t}(|r_{1}r_{2} \rangle\langle r_{1}g_{2}|+e^{i\theta_{1}}|r_{1}r_{2} \rangle\langle g_{1}r_{2}|)+\text{H.c.}.
\label{HA_x}
\end{split}
\end{equation}
Here, $a_{L}=a_{L}^{\prime}e^{-ik_{0}x}$ ($a_{L}^{\dagger}=a_{L}^{\prime\dagger}e^{ik_{0}x}$) and $a_{R}=a_{R}^{\prime}e^{ik_{0}x}$ ($a_{R}^{\dagger}=a_{R}^{\prime\dagger}e^{-ik_{0}x}$) denote the annihilation (creation) operators of left-going and right-going photons in waveguide mode $a$, respectively, with $a_{L}^{\prime}$ and $a_{R}^{\prime}$ representing spatially slowly-varying envelopes. Note also that the delta functions $\delta(x)$ and $\delta(x-d)$ describe two spatially separated coupling points, and we have chosen $\omega_{0}=\omega_{e}$ as the frequency around which the dispersion relation of waveguide mode $a$ can be linearized as $\omega_{ka}=\omega_{0}-(k_{0}\pm k_{a})v_{g}$ with `$+$' (`$-$') referring to the left (right) branch.
For a waveguide photon of central frequency $\omega_{ka}$ (its detuning is defined as $\delta_{ka}=\omega_{ka}-\omega_{e}$) incident from port 1 on the left or from port 2 on the right, one can solve the stationary Schr\"{o}dinger equation
$H_{A_{x}}|\Psi_{A_{x}}\rangle=\omega_{ka}|\Psi_{A_{x}}\rangle$ to examine the following eigenstate
\begin{equation}
\begin{split}
|\Psi_{A_{x}}\rangle&=\int_{-\infty}^{+\infty}dx\big\lbrack\Phi_{aR}^{A}(x)a_{R}^{\dagger}(x)+\Phi_{aL}^{A}(x)a_{L}^{\dagger}(x)\big\rbrack|0,g_{1}g_{2}\rangle\\
&+u_{b}^{A}|0,r_{1}g_{2}\rangle+u_{c}^{A}|0,g_{1}r_{2}\rangle+u_{d}^{A}|0,r_{1}r_{2}\rangle,
\label{Phi_A}
\end{split}
\end{equation}
where $|0, g_{1}g_{2}\rangle$ denotes the vacuum state of the system without waveguide photons and excited atoms;
$u^{A}_{b}$, $u^{A}_{c}$, and $u^{A}_{d}$ are the excitation probability amplitudes of different atomic states. Note that all terms in $|\Psi_{A_{x}}\rangle$ belong to the single-excitation subspace of waveguide photons, as the double-excitation state $|r_{1}r_{2}\rangle$ corresponds to a two-photon transition from the ground state $|g_{1}g_{2}\rangle$ in which only one quantum of the waveguide mode is exchanged (the other comes from the driving field).

Moreover, under appropriate boundary conditions, the densities of probability amplitudes for right-going and left-going photons can be written as
\begin{equation}
\begin{split}
\Phi_{aR}^{A}(x)&=e^{ik_{a}x}\big\lbrace\Theta(-x)+A_{1}\big\lbrack\Theta(x)-\Theta(x-d
)\big\rbrack\\
&+t_{1\rightarrow2}\Theta(x-d)\big\rbrace,\\
\Phi_{aL}^{A}(x)&=e^{-ik_{a}x}\big\lbrace r_{1\rightarrow1}\Theta(-x)+A_{2}\big\lbrack\Theta(x)-\Theta(x-d
)\big\rbrack\big\rbrace,
\label{Phi_a_rl}
\end{split}
\end{equation}
referring to the case where one photon is incident from the left (i.e., port 1) of the waveguide, or
\begin{equation}
\begin{split}
\Phi_{aR}^{A}(x)&=e^{ik_{a}x}\big\lbrace A_{3}\big\lbrack\Theta(x)-\Theta(x-d
)\big\rbrack+r_{2\rightarrow2}\Theta(x-d)\big\rbrace,\\
\Phi_{aL}^{A}(x)&=e^{-ik_{a}x}\big\lbrace\Theta(x-d)+A_{4}\big\lbrack\Theta(x)-\Theta(x-d
)\big\rbrack\\
&+t_{2\rightarrow1}\Theta(-x)\big\rbrace,
\label{Phi_a_rl2}
\end{split}
\end{equation}
referring to the case where one photon is incident from the right (i.e., port 2) of the waveguide. All unknown coefficients appearing in $|\Psi_{A_{x}}\rangle$ can be obtained straightforwardly (see Supplementary IA) from the solution of $H_{A_{x}}|\Psi_{A_{x}}\rangle=\omega_{ka}|\Psi_{A_{x}}\rangle$, This then yields two transmissivities $T_{1\rightarrow 2}=|t_{1\rightarrow 2}|^2$ and $T_{2\rightarrow 1}=|t_{2\rightarrow 1}|^2$, which are in general different (i.e., nonreciprocal) due to equivalent giant-atom chiral couplings as discussed below.

\subsection{Equivalent giant atom}

Under the two-photon resonance condition (i.e., $\Delta_{c1}+\delta_{ka}\simeq0$) and with detunings much larger than coupling strengths (i.e., $|\delta_{ka}|\simeq|\Delta_{c1}|\gg\Omega_{c1},g_{a}$), it has been shown that the double-excitation state $|r_{1}r_{2}\rangle$ decays directly to the ground state $|g_{1}g_{2}\rangle$, exhibiting a giant-atom self-interference behavior~\cite{RydGiant}. Now we try to show that the scattering properties of two Rydberg atoms can also be equivalent to those of a giant atom under the same considerations. For this purpose, we assume that both atoms are initially at state $|g_{1}g_{2}\rangle$ and will be excited to state $|r_{1}r_{2}\rangle$ directly by a waveguide-mode photon and a driving-field photon together, leaving the single-excitation states $|r_{1}g_{2}\rangle$ and $|g_{1}r_{2}\rangle$ almost unpopulated during the scattering process. Then, one can adiabatically eliminate states $|r_{1}g_{2}\rangle$ and $|g_{1}r_{2}\rangle$ \cite{elimination1,elimination2} and obtain the \textit{effective} real-space Hamiltonian of a two-level giant-atom as shown in Fig.~\figpanel{fig1}{c} from its momentum-space representation (see Supplementary IB) as \cite{fan2}
\begin{equation}
\begin{split}
H_{A_{x}}^{\text{eff}}&=(2\omega_{e}+V_{6}-\frac{2\Omega_{c1}^{2}}{\Delta_{c1}}-2i\gamma)|r_{1}r_{2} \rangle\langle r_{1}r_{2}|\\
&+\int_{-\infty}^{+\infty}dx\Big\lbrack a_{L}^{\prime\dagger}(x)\Big(\omega_{c1}+\omega_{0}+iv_{g}\frac{\partial} {\partial x}\Big)a_{L}^{\prime}(x)\\
&+a_{R}^{\prime\dagger}(x)\Big(\omega_{c1}+\omega_{0}-iv_{g}\frac{\partial} {\partial x}\Big)a_{R}^{\prime}(x)\Big\rbrack\\
&+\Big\lbrace\int_{-\infty}^{+\infty}dx\xi_{a}\left[a_{R}(x)+a_{L}(x)\right]|r_{1}r_{2} \rangle\langle g_{1}g_{2}|\\
&\times\big[\delta(x)+\delta(x-d)e^{i\theta_{1}}\big]+\text{H.c.}\Big\rbrace,
\label{HA_x^eff}
\end{split}
\end{equation}
where $2\Omega_{c1}^{2}/\Delta_{c1}$ is the effective energy shift of state $|r_{1}r_{2}\rangle$ and $\xi_{a}=-g_{a}\Omega_{c1}/\Delta_{c1}$ is the effective coupling strength of relevant two-photon transition processes.

\begin{figure}[ptb]
\centering
\includegraphics[width=8.5 cm]{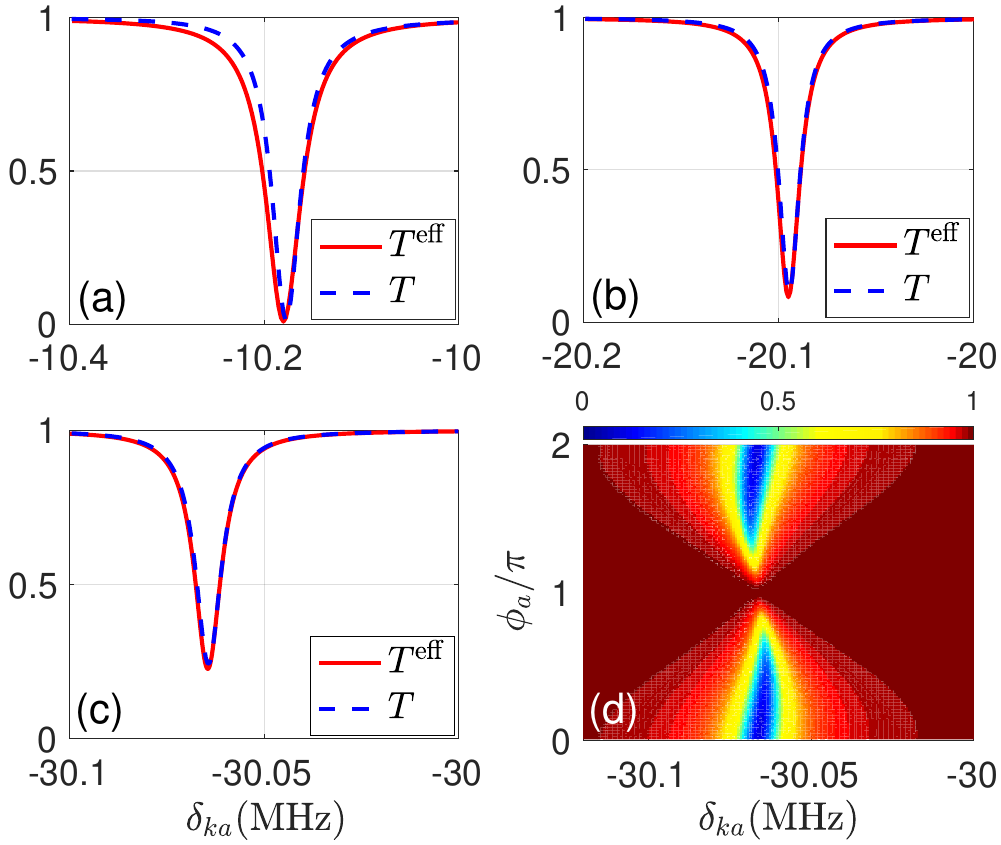}
\caption{Reciprocal transmissivities $T$ and $T^{\text{eff}}$ versus detuning $\delta_{ka}$ with $\phi_{a}=\pi/2$ and (a) $\Delta_{c1}=10$ MHz; (b) $\Delta_{c1}=20$ MHz; (c) $\Delta_{c1}=30$ MHz. (d) Reciprocal transmissivity $T$ ($T^{\text{eff}}$) versus detuning $\delta_{ka}$ and phase $\phi_{a}$ with $\Delta_{c1}=30$ MHz. Other parameters are taken as $\theta_{1}=0$, $\gamma=1$ kHz, $\Gamma_{a}=1$ MHz, $\Omega_{c1}=1$ MHz, and $V_{6}=20$ GHz.}
\label{fig2}
\end{figure}

Again, the eigenstate of $H_{A_{x}}^{\text{eff}}|\tilde{\Psi}_{A_{x}}\rangle=(\omega_{ka}+\omega_{c1})|\tilde{\Psi}_{A_{x}}\rangle$ can be written as (Bethe-ansatz approach)
\begin{equation}
\begin{split}
|\tilde{\Psi}_{A_{x}}\rangle&=\int_{-\infty}^{+\infty}dx\big\lbrack\tilde{\Phi}^{A}_{aR}(x)a_{R}^{\dagger}(x)+\tilde{\Phi}^{A}_{aL}(x)a_{L}^{\dagger}(x)\big\rbrack|0,g_{1}g_{2}\rangle\\
&+\tilde{u}^{A}_{d}|0,r_{1}r_{2}\rangle,
\label{Phi_A^eff}
\end{split}
\end{equation}
where $\tilde{\Phi}^{A}_{aL}$, $\tilde{\Phi}^{A}_{aR}$, and $\tilde{u}^{A}_{d}$ have similar physical meanings as $\Phi^{A}_{aL}$, $\Phi^{A}_{aR}$, and $u^{A}_{d}$ in Eq~(\ref{Phi_A}), respectively. Then, for a photon of central frequency $\omega_{ka}$ incident from port 1 or port 2 of the waveguide, one can obtain the analytical expressions of transmissivities $T_{1\rightarrow2}^{\text{eff}}=|t_{1\rightarrow2}^{\text{eff}}|^{2}$ and $T_{2\rightarrow1}^{\text{eff}}=|t_{2\rightarrow1}^{\text{eff}}|^{2}$ (see Supplementary IB) as
\begin{equation}
\begin{split}
T_{1\rightarrow2}^{\text{eff}}&=\left|\frac{\delta_{ka}+\Delta_{c1}+\frac{2\Omega_{c1}^{2}}{\Delta_{c1}}+2i\gamma-2\Upsilon_{a}e^{i\theta_{1}}\text{sin}\phi_{a}}
{\delta_{ka}+\Delta_{c1}+\frac{2\Omega_{c1}^{2}}{\Delta_{c1}}+2i\gamma+2i\Upsilon_{a}(1+ e^{i\phi_{a}}\text{cos}\theta_{1})}\right|^{2},\\
T_{2\rightarrow1}^{\text{eff}}&=\left|\frac{\delta_{ka}+\Delta_{c1}+\frac{2\Omega_{c1}^{2}}{\Delta_{c1}}+2i\gamma-2\Upsilon_{a}e^{-i\theta_{1}}\text{sin}\phi_{a}}
{\delta_{ka}+\Delta_{c1}+\frac{2\Omega_{c1}^{2}}{\Delta_{c1}}+2i\gamma+2i\Upsilon_{a}(1+ e^{i\phi_{a}}\text{cos}\theta_{1})}\right|^{2},
\label{t^eff}
\end{split}
\end{equation}
with $\Upsilon_{a}=\xi_{a}^2/v_{g}$ denoting the effective decay rate into waveguide mode $a$ and $\phi_{a}=k_{a}d=k_{0}d+(\omega_{ka}-\omega_{0})d/v_{g}$ that can be taken as constant in the Markovian regime \cite{cp2022}. In this subsection, we wish to test the validity of adiabatically eliminating two single-excitation states, and first focus on the case of a normal driving field's incidence ($\theta_{1}=0$) corresponding to a reciprocal transmission with $T^{\text{eff}}=T_{1\rightarrow2}^{\text{eff}}=T_{2\rightarrow1}^{\text{eff}}$ and $T=T_{1\rightarrow 2}=T_{2\rightarrow 1}$ (\textit{nonchiral case}).

Comparing the spectra of transmissivities $T$ from the four-level double-atom model and $T^{\text{eff}}$ from the two-level giant-atom model in Figs.~\figpanel{fig2}{a}-\figpanel{fig2}{c} for different values of $\Delta_{c1}$, it is clear that $T$ and $T^{\text{eff}}$ become closer and closer as $\Delta_{c1}$ increases, indicating that the adiabatic elimination becomes more and more reliable. At the same time, we note that the effective decay rate $\Upsilon_{a}=\Gamma_{a}\Omega_{c1}^{2}/\Delta_{c1}^{2}$ becomes smaller for larger $\Delta_{c1}$, which makes the minimum values of $T$ and $T^{\text{eff}}$, manifested as a dip around $\delta_{ka}\simeq-\Delta_{c1}$, to increase. In particular, we find that the two-level giant-atom model is very accurate at $\Delta_{c1}=30$ MHz, for which we have plotted $T$ ($T^{\text{eff}}$) against detuning $\delta_{ka}$ and phase $\phi_{a}$ in Fig.~\figpanel{fig2}{d}. This plot shows the typical phase-dependent transmission spectra of a giant atom due to the self-interference effect between two coupling points. It is also worth noting that $T^{\text{eff}} (T)\equiv 1$ with $\phi_{a}=(2n+1)\pi$ ($n$ is an integer) refers to the specific case of two Rydberg atoms equivalent to a giant atom being decoupled from the waveguide mode. All above findings can find appropriate explanations from Eq.~(\ref{t^eff}).

\begin{figure}[ptb]
\centering
\includegraphics[width=8.5 cm]{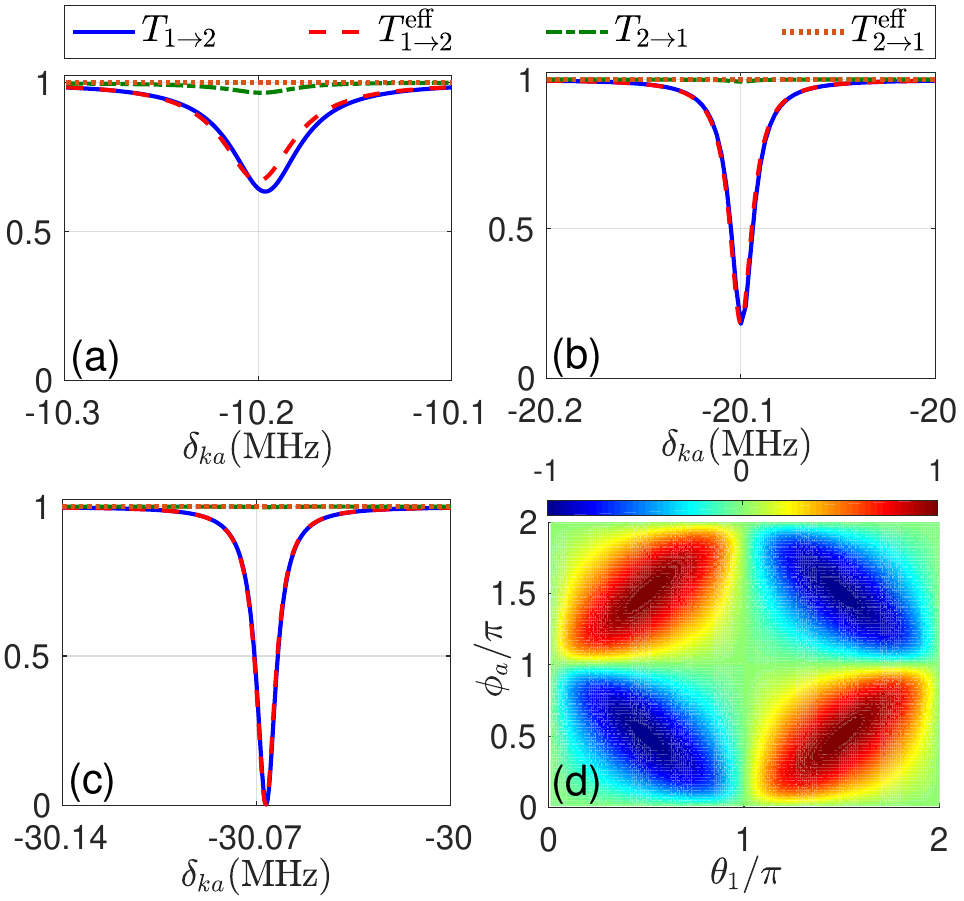}
\caption{Nonreciprocal transmissivities $T_{1\rightarrow2}$ and $T_{2\rightarrow1}$ as well as $T_{1\rightarrow2}^{\text{eff}}$ and $T_{2\rightarrow1}^{\text{eff}}$ versus detuning $\delta_{ka}$ with $\phi_{a}=\theta_{1}=\pi/2$ and (a) $\Delta_{c1}=10$ MHz; (b) $\Delta_{c1}=20$ MHz; (c) $\Delta_{c1}=30$ MHz. (d) Transmission contrast ratio $I$ versus phases $\theta_{1}$ and $\phi_{a}$ with $\delta_{ka}=-30.067$ MHz and $\Delta_{c1}=30$ MHz. Other parameters are the same as in Fig.~\ref{fig2}.}
\label{fig3}
\end{figure}

\subsection{Equivalent chiral couplings}

It is now worth recalling that coupling phases may be introduced, e.g., through Josephson-junction loops when threaded by external fluxes~\cite{WXarxiv} or through a dissipation port at the coupling point~\cite{prx2023} in superconducting quantum systems.
Here, we present a much simpler method to realize chiral couplings in the optical domain based on our Rydberg giant-atom platform, where the chirality is attained by controlling the non-vanishing local phase difference $\theta_{1}$ between two coupling points of the equivalent giant atom~\cite{cp2022,WXarxiv,DLprl}. Directly adjusting the incident angle $\alpha$ of an oblique driving field would then enable one, in the presence of intrinsic atomic dissipation ($\gamma\ne0$), to realize the nonreciprocal transmission.

\begin{figure}[ptb]
\centering
\includegraphics[width=8.5 cm]{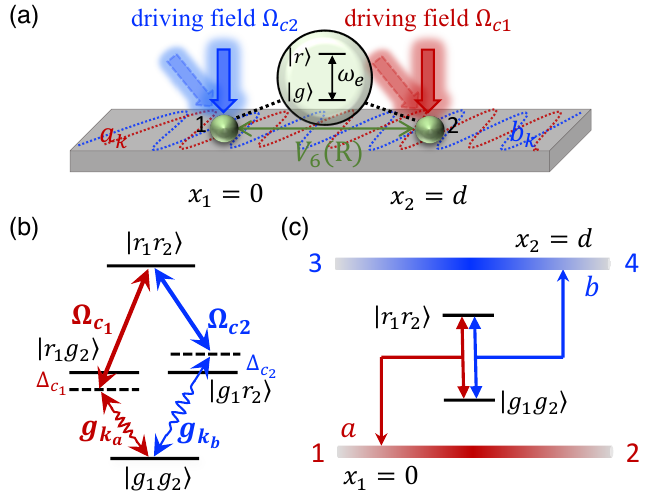}
\caption{Schematic diagram for achieving symmetric frequency conversion between two waveguide modes. (a) The same waveguide-coupled Rydberg-atom platform as in Fig.~\figpanel{fig1}{a} except atoms $1$ and $2$ are driven by coherent fields $\Omega_{c1}$ and $\Omega_{c2}$ of different incident angles, respectively. (b) The four-level configuration in the two-atom basis where driving field $\Omega_{c1}$ ($\Omega_{c2}$) is matched in frequency with waveguide mode $a$ ($b$) on the left (right) path of two-photon resonance. (c) The equivalent two-level nonlocal atom coupled to each waveguide mode at one point when eliminating two single-excitation states.} \label{fig4}
\end{figure}

We plot in Figs.~\figpanel{fig3}{a}-\figpanel{fig3}{c} relevant transmissivities against detuning $\delta_{ka}$ when a waveguide photon is incident from port 1 on the left or port 2 on the right in the case of perfect chiral couplings with $\phi_{a}=\pi/2$ and $\theta_{1}=\pi/2$. It is clear that, around the two-photon resonance $\delta_{ka}\simeq-\Delta_{c1}$, transmissivities $T_{1\rightarrow2}$ and $ T_{1\rightarrow2}^{\text{eff}}$ from port 1 to port 2 (or $T_{2\rightarrow1}$ and $T_{2\rightarrow1}^{\text{eff}}$ from port 2 to port 1) are very different for different values of $\Delta_{c1}$. That is because, other than determining the validity of adiabatically eliminating two single-excitation states [the same conclusion is depicted in Figs.~\figpanel{fig2}{a}-\figpanel{fig2}{c}], a change of $\Delta_{c1}$ will result in different values of the waveguide decay rate $\Gamma_{a}$ while the intrinsic dissipation rate $\gamma$ remains unchanged. For instance, in Fig.~\figpanel{fig3}{c} we have $T_{1\rightarrow2}\simeq T_{1\rightarrow2}^{\text{eff}}\simeq0$ ($T_{2\rightarrow1}\simeq T_{2\rightarrow1}^{\text{eff}}\equiv1$ ) at the resonance point when $\Upsilon_{a}=\Gamma_{a}\Omega_{c1}^{2}/\Delta_{c1}^{2}$ corresponding to $\Delta_{c1}=30$ MHz and $\gamma$ are almost the same to result in the perfect destructive (constructive) interference for a right-going (left-going) waveguide photon as can be seen from Eq.~(\ref{t^eff}). Therefore, with the experimentally reasonable parameters of $\Upsilon_{a}$ here, Rydberg atoms are the most suitable to realize the perfect nonreciprocal transmission due to the fact that their intrinsic dissipations are small compared with normal natural atoms. Furthermore, the transmission contrast ratio
\begin{equation}
I=\frac{T_{2\rightarrow1}-T_{1\rightarrow2}}{T_{2\rightarrow1}+T_{1\rightarrow2}},
\label{I}
\end{equation}
plotted in Fig.~\figpanel{fig3}{d} shows that the transmission nonreciprocity due to chiral couplings can be easily tuned by relative phases $\theta_{1}$ and $\phi_{a}$. Note also that $\gamma$ plays a crucial role here and we have $I\equiv0$ when neglecting $\gamma$ \cite{cp2022}.

Since the spatial extents of highly-excited Rydberg states may compare with the optical wavelengths of driving fields and waveguide modes, the issue of continuous couplings, rather than point-like couplings, should be considered. But, we can verify through a dynamical solution method in the momentum space (see Supplementary IC) that the nonreciprocal transmission is robust since all results found above remain essentially quantitatively unchanged even for continuous couplings.

\section{Symmetric frequency conversion}

In addition to realizing nonreciprocal transmission for one waveguide mode, another possible application of our Rydberg-atom platform is to attain frequency conversion between two waveguide modes. This can be implemented by applying coherent fields $\Omega_{c1}$ and $\Omega_{c2}$ with frequencies $\omega_{c1}$ and $\omega_{c2}$ upon atoms $1$ and $2$, respectively, as shown in Fig.~\figpanel{fig4}{a}. If we further assume $|\omega_{c1}-\omega_{c2}|\gg\Omega_{c1,c2}$, each atom will couple to a different waveguide mode ($a$ or $b$), rather than the same waveguide mode ($a$) as shown in Fig.~\figpanel{fig1}{a}, under the two-photon resonance from ground state $|g_{1}g_{2}\rangle$ to double-excitation state $|r_{1}r_{2}\rangle$. This then results in the four-level configuration in the two-atom basis as shown in Fig.~\figpanel{fig4}{b}, whereby driving field $\Omega_{c1}$ and waveguide mode $a$ form the left path while driving field $\Omega_{c2}$ and waveguide mode $b$ form the right path. Accordingly, the system Hamiltonian can be written as
\begin{equation}
\begin{split}
H_{B_{k}}&=(\omega_{e}-i\gamma)(|g_{1}r_{2}\rangle\langle g_{1}r_{2}|+|r_{1}g_{2}\rangle\langle r_{1}g_{2}|)\\
&+(2\omega_{e}+V_{6}-2i\gamma)|r_{1}r_{2}\rangle\langle r_{1}r_{2}|+\int dk_{a}\omega_{ka}a_{k}^{\dagger}a_{k}\\
&+\int dk_{b}\omega_{kb}b_{k}^{\dagger}b_{k}+\Big[\int dk_{a}g_{a}a_{k}|r_{1}g_{2} \rangle\langle g_{1}g_{2}|\\
&+\int dk_{b}g_{b}b_{k}|g_{1}r_{2} \rangle\langle g_{1}g_{2}|e^{ik_{b}d}\\
&+\Omega_{c1} e^{-i\omega_{c1}t}|r_{1}r_{2} \rangle\langle r_{1}g_{2}|\\
&+\Omega_{c2} e^{-i\omega_{c2}t}e^{i\theta_{2}}|r_{1}r_{2} \rangle\langle g_{1}r_{2}|+\text{H.c.}\Big],
\label{HB_x}
\end{split}
\end{equation}
where $b_{k}^{\dagger}$ ($b_{k}$) is the bosonic creation (annihilation) operators of the second waveguide mode with frequency $\omega_{kb}$ and wave vector $k_{b}$; $g_{b}$ is the constant coupling strength between atom $2$ and waveguide mode $b$ and we will assume $g_{b}=g_{a}$ for simplicity; $\theta_{2}=k_{c2}l_{2}$ is the phase difference with respect to optical path difference $l_{2}=d\sin\beta$ for the second driving field at incident angle $\beta$. Transferring $H_{B_{k}}$ in the momentum space into $H_{B_{x}}$ in the real space, one can calculate scattering possibilities and conversion efficiencies of a right-going or left-going photon in waveguide mode $a$ by solving the stationary Schr\"{o}dinger equation as done in the last section for $H_{A_{x}}$. 

\begin{figure}[ptb]
\centering
\includegraphics[width=8 cm]{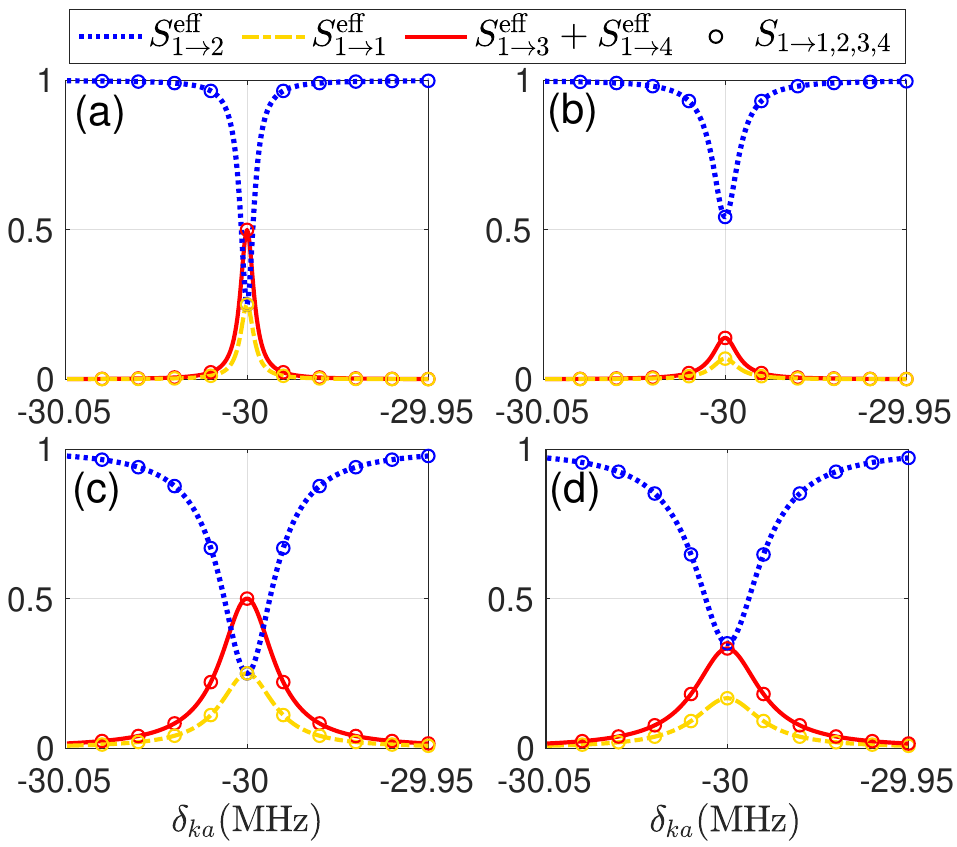}
\caption{Effective reflectivity $S_{1\rightarrow1}^{\rm{eff}}$, transmissivity $S_{1\rightarrow2}^{\rm{eff}}$, and total conversion efficiency $S_{1\rightarrow3}^{\rm{eff}}+S_{1\rightarrow4}^{\rm{eff}}$ (compared with $S_{1\rightarrow1}$, $S_{1\rightarrow2}$, and $S_{1\rightarrow3}+S_{1\rightarrow4}$) versus detuning $\delta_{ka}$ with (a) $\gamma=0$ and $\Omega_{c1}=\Omega_{c2}=1$ MHz; (b) $\gamma=1$ kHz and $\Omega_{c1}=\Omega_{c2}=1$ MHz; (c) $\gamma=0$ and $\Omega_{c1}=\Omega_{c2}=2$ MHz; (d) $\gamma=1$ kHz and $\Omega_{c1}=\Omega_{c2}=2$ MHz. Other parameters are $\Gamma_{a}=\Gamma_{b}=1$ MHz, $\Delta_{c1}=-\Delta_{c2}=30$ MHz, and $V_{6}=20$ GHz.}
\label{fig5}
\end{figure}

In a similar way, adiabatically eliminating the single-excitation states $|r_{1}g_{2}\rangle$ and $|g_{1}r_{2}\rangle$, one can obtain the effective real-space Hamiltonian from its momentum space representation (see Supplementary II) as
\begin{equation}
\begin{split}
H_{B_{x}}^{\text{eff}}&=(2\omega_{e}+V_{6}-\frac{\Omega_{c1}^{2}}{\Delta_{c1}}-\frac{\Omega_{c2}^{2}}{\Delta_{c2}}-2i\gamma)|r_{1}r_{2} \rangle\langle r_{1}r_{2}|\\
&+\int_{-\infty}^{+\infty}dx\Big\lbrack a_{L}^{\prime\dagger}(x)\Big(\omega_{c1}+\omega_{0}+iv_{g}\frac{\partial} {\partial x}\Big)a_{L}^{\prime}(x)\\
&+a_{R}^{\prime\dagger}(x)\Big(\omega_{c1}+\omega_{0}-iv_{g}\frac{\partial} {\partial x}\Big)a_{R}^{\prime}(x)\Big\rbrack\\
&+\int_{-\infty}^{+\infty}dx\Big\lbrack b_{L}^{\prime\dagger}(x)\Big(\omega_{c2}+\omega_{0}+iv_{g}\frac{\partial} {\partial x}\Big)b_{L}^{\prime}(x)\\
&+b_{R}^{\prime\dagger}(x)\Big(\omega_{c2}+\omega_{0}-iv_{g}\frac{\partial} {\partial x}\Big)b_{R}^{\prime}(x)\Big\rbrack\\
&+\int_{-\infty}^{+\infty}dx\big\lbrace\xi_{a}\left[a_{R}(x)+a_{L}(x)\right]|r_{1}r_{2} \rangle\langle g_{1}g_{2}|\delta(x)\\
&+\xi_{b}\left[b_{R}(x)+b_{L}(x)\right]|r_{1}r_{2} \rangle\langle g_{1}g_{2}|\delta(x-d)e^{i\theta_{2}}+\text{H.c.}\big\rbrace,
\label{HB_x^eff}
\end{split}
\end{equation}
with $\Delta_{c2}=\omega_{c2}-(\omega_{e}+V_{6})$ and $\xi_{b}=-g_{b}\Omega_{c2}/\Delta_{c2}$. Here, $b_{L}=b_{L}^{\prime}e^{-ik_{0}x}$ ($b_{L}^{\dagger}=b_{L}^{\prime\dagger}e^{ik_{0}x}$) and $b_{R}=b_{R}^{\prime}e^{ik_{0}x}$ ($b_{R}^{\dagger}=b_{R}^{\prime\dagger}e^{-ik_{0}x}$) denote the annihilation (creation) operators of left-going and right-going photons in waveguide mode $b$, respectively. This Hamiltonian refers to an effective two-level system as shown in Fig.~\figpanel{fig4}{c} that should be regarded as a nonlocal atom (but not a giant atom) since each of the two involved waveguide modes exhibits a single coupling point, thus yielding no self-interference effect. Its eigenstate can be written as
\begin{equation}
\begin{split}
|\tilde{\Psi}_{B_{x}}\rangle&=\int_{-\infty}^{+\infty}dx\big\lbrack\tilde{\Phi}^{B}_{aR}(x)a_{R}^{\dagger}(x)+\tilde{\Phi}^{B}_{aL}(x)a_{L}^{\dagger}(x)\\
&+\tilde{\Phi}^{B}_{bR}(x)b_{R}^{\dagger}(x)+\tilde{\Phi}^{B}_{bL}(x)b_{L}^{\dagger}(x)\big\rbrack|0,g_{1}g_{2}\rangle\\
&+\tilde{u}^{B}_{d}|0,r_{1}r_{2}\rangle,
\label{Phi_B^eff}
\end{split}
\end{equation}
where $\tilde{\Phi}^{B}_{aR,aL}(x)$ and $\tilde{\Phi}^{B}_{bR,bL}(x)$ describe the densities of probability amplitudes for a right-going or left-going photon in waveguide modes $a$ and $b$, respectively.

In this case, for example, if a single photon is incident from port 1 or port 2 of waveguide mode $a$, two ground-state Rydberg atoms will make a two-photon transition to the double-excitation state, simultaneously extracting another photon from the first driving field $\Omega_{c1}$. Then, they will relax back to the ground state, emitting a photon into the other waveguide mode $b$ along with another photon into the second driving field $\Omega_{c2}$, or a photon into the original waveguide mode $a$ along with another photon into the first driving field $\Omega_{c1}$. Based on the Bethe-ansatz approach, one has (see Supplementary II)
\begin{equation}
\begin{split}
S_{1\rightarrow1}^{\text{eff}}=|s_{1\rightarrow1}^{\text{eff}}|^2&=\left|\frac{-i\Upsilon_{a}}{i(\Upsilon_{a}+\Upsilon_{b})+(\delta_{ka}+\Delta_{c1}+2i\gamma)}\right|^{2},\\
S_{1\rightarrow2}^{\text{eff}}=|s_{1\rightarrow2}^{\text{eff}}|^2&=\left|\frac{i\Upsilon_{b}+(\delta_{ka}+\Delta_{c1}+2i\gamma)}{i(\Upsilon_{a}+\Upsilon_{b})+(\delta_{ka}+\Delta_{c1}+2i\gamma)}\right|^{2},\\
S_{1\rightarrow3}^{\text{eff}}=|s_{1\rightarrow3}^{\text{eff}}|^2&=\left|\frac{-i\sqrt{\Upsilon_{a}\Upsilon_{b}}e^{i(\theta_{2}+\phi_{b})}}{i(\Upsilon_{a}+\Upsilon_{b})+(\delta_{ka}+\Delta_{c1}+2i\gamma)}\right|^{2},\\
S_{1\rightarrow4}^{\text{eff}}=|s_{1\rightarrow4}^{\text{eff}}|^2&=\left|\frac{-i\sqrt{\Upsilon_{a}\Upsilon_{b}}e^{i(\theta_{2}-\phi_{b})}}{i(\Upsilon_{a}+\Upsilon_{b})+(\delta_{ka}+\Delta_{c1}+2i\gamma)}\right|^{2},
\label{S^eff}
\end{split}
\end{equation}
with $\Upsilon_{b}=\xi_{b}^{2}/v_{g}=\Gamma_{b}\Omega_{c2}^{2}/\Delta_{c2}^{2}$ and $\phi_{b}=k_{b}d$. We have also set $\delta_{kb}=\omega_{kb}-\omega_{e}=-\delta_{ka}$ due to the requirement of energy conservation in a close-loop giant-atom transition with $\Delta_{c2}=-\Delta_{c1}$. Here, $S_{1\rightarrow1}^{\text{eff}}$ ($S_{1\rightarrow2}^{\text{eff}}$) denotes the effective reflectivity (transmissivity) of mode $a$, while $S_{1\rightarrow3}^{\text{eff}}$ ($S_{1\rightarrow4}^{\text{eff}}$) refers to the effective backward (forward) conversion efficiencies into waveguide mode $b$. Note that the right-incident case (i.e., from port 2) is a symmetric process compared with the left-incident one (i.e., from port 1) as considered above because $S_{1\rightarrow 1,2,3,4}^{\text{eff}}$ are reciprocal, being insensitive to phases $\theta_2$ and $\phi_b$ even if the two driving fields are oblique (i.e., $\alpha\ne0$ and $\beta\ne0$).

We plot in Fig.~\ref{fig5} $S_{1\rightarrow 1}^{\text{eff}}$, $S_{1\rightarrow 2}^{\text{eff}}$, and $S_{1\rightarrow 3}^{\text{eff}}+S_{1\rightarrow 4}^{\text{eff}}$  by comparing them with $S_{1\rightarrow 1}$, $S_{1\rightarrow 2}$, and $S_{1\rightarrow 3}+S_{1\rightarrow 4}$ obtained from the original Hamiltonian $H_{B_{x}}$  with $\Delta_{c1}=-\Delta_{c2}=30$ MHz. It is evident that the adiabatic elimination of two single-excitation states is valid again since there is no difference between $S_{1\rightarrow j}^{\text{eff}}$ and $S_{1\rightarrow j}$. Moreover, the optimal value of frequency conversion efficiency quantified by $S_{1\rightarrow3}^{\text{eff}}+S_{1\rightarrow4}^{\text{eff}}$ is only $0.5$ when ignoring the intrinsic atomic dissipation ($\gamma=0$) and taking $\Upsilon_{a}=\Upsilon_{b}$, as shown in Figs.~\figpanel{fig5}{a} and~\figpanel{fig5}{c}. In fact, we have $S_{1\rightarrow3}^{\text{eff}}\equiv S_{1\rightarrow4}^{\text{eff}}$ and they are further equal to $S_{1\rightarrow2}^{\text{eff}}$ if we set $\Upsilon_{a}=\Upsilon_{b}$ as can be seen from Eq.~(\ref{S^eff}). The inclusion of $\gamma$ in Figs.~\figpanel{fig5}{b} and~\figpanel{fig5}{d} would lower the conversion efficiency, yet one could increase Rabi frequencies $\Omega_{c1}$ and $\Omega_{c2}$ to raise $\Upsilon_{a}=\Upsilon_{b}$, thus reducing the negative effect of $\gamma$. This prompts us to investigate another scenario for improving the conversion efficiency and meanwhile achieving the asymmetric scattering.

\section{Asymmetric frequency conversion}

\begin{figure}[pth]
\centering
\includegraphics[width=8.5 cm]{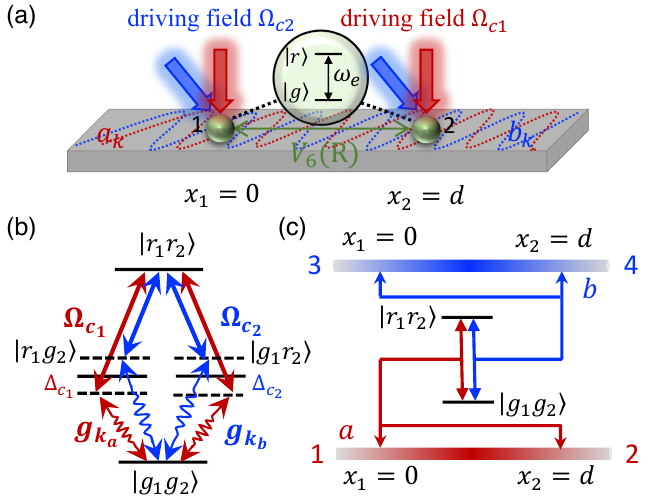}
\caption{Schematic diagram for achieving asymmetric frequency conversion between two waveguide modes. (a) The same waveguide-coupled Rydberg-atom platform as in Fig.~\figpanel{fig4}{a} except atoms $1$ and $2$ are both driven by coherent fields $\Omega_{c1}$ and $\Omega_{c2}$ at different angles of incidence. (b) The four-level configuration in the two-atom basis where driving field $\Omega_{c1}$ ($\Omega_{c2}$) is matched in frequency with waveguide mode $a$ ($b$) on both left and right paths of two-photon resonance. (c) The equivalent two-level giant atom coupled to each waveguide mode at two points when eliminating two single-excitation states.}\label{fig6}
\end{figure}

We further consider the case where each Rydberg atom is driven by two coherent fields $\Omega_{c1}$ and $\Omega_{c2}$ as shown in Fig.~\figpanel{fig6}{a}. This then results in the four-level configuration in the two-atom basis as shown in Fig.~\figpanel{fig6}{b}, where both left and right paths of the two photon resonant transition from ground state $|g_{1}g_{2}\rangle$ to double-excitation state $|r_{1}r_{2}\rangle$ can be implemented with driving field $\Omega_{c1}$ and waveguide mode $a$ or with driving field $\Omega_{c2}$ and waveguide mode $b$. As discussed below, this means of all-optical control will enable one to enhance the efficiency and select the directionality of frequency conversion via tunable chiral couplings, yet without altering the system's physical structure. The relevant Hamiltonian reads as
\begin{equation}
\begin{split}
H_{C_{k}}&=(\omega_{e}-i\gamma)(|g_{1}r_{2}\rangle\langle g_{1}r_{2}|+|r_{1}g_{2}\rangle\langle r_{1}g_{2}|)\\
&+(2\omega_{e}+V_{6}-2i\gamma)|r_{1}r_{2}\rangle\langle r_{1}r_{2}|\\
&+\int dk_{a}\omega_{ka}a_{k}^{\dagger}a_{k}+\int dk_{b}\omega_{kb}b_{k}^{\dagger}b_{k}\\
&+\Big[\int dk_{a}g_{a}a_{k}(|r_{1}g_{2} \rangle\langle g_{1}g_{2}|+|g_{1}r_{2} \rangle\langle g_{1}g_{2}|e^{ik_{a}d})\\
&+\int dk_{b}g_{b}b_{k}(|r_{1}g_{2} \rangle\langle g_{1}g_{2}|+|g_{1}r_{2} \rangle\langle g_{1}g_{2}|e^{ik_{b}d})\\
&+\Omega_{c1} e^{-i\omega_{c1}t}(|r_{1}r_{2} \rangle\langle r_{1}g_{2}|+e^{i\theta_{1}}|r_{1}r_{2} \rangle\langle g_{1}r_{2}|)\\
&+\Omega_{c2} e^{-i\omega_{c2}t}(|r_{1}r_{2} \rangle\langle r_{1}g_{2}|+e^{i\theta_{2}}|r_{1}r_{2} \rangle\langle g_{1}r_{2}|)+\text{H.c.}\Big],
\label{HC_k}
\end{split}
\end{equation}
where the two local phase differences $\theta_{1}$ and $\theta_{2}$ as defined before can be independently controlled by changing the respective driving fields' angles of incidence.

After adiabatically eliminating the single-excitation states $|r_{1}g_{2}\rangle$ and $|g_{1}r_{2}\rangle$ in a way similar to that considered in the last two sections, one can attain from $H_{C_{k}}$ the effective Hamiltonian $H_{C_{k}}^{\text{eff}}$ (see Supplementary III), which if transferred into the real space becomes
\begin{equation}
\begin{split}
H_{C_{x}}^{\text{eff}}&=(2\omega_{e}+V_{6}-\frac{2\Omega_{c1}^{2}}{\Delta_{c1}}-\frac{2\Omega_{c2}^{2}}{\Delta_{c2}}-2i\gamma)|r_{1}r_{2} \rangle\langle r_{1}r_{2}|\\
&+\int_{-\infty}^{+\infty}dx\Big\lbrack a_{L}^{\prime\dagger}(x)\Big(\omega_{c1}+\omega_{0}+iv_{g}\frac{\partial} {\partial x}\Big)a_{L}^{\prime}(x)\\
&+a_{R}^{\prime\dagger}(x)\Big(\omega_{c1}+\omega_{0}-iv_{g}\frac{\partial} {\partial x}\Big)a_{R}^{\prime}(x)\Big\rbrack\\
&+\int_{-\infty}^{+\infty}dx\Big\lbrack b_{L}^{\prime\dagger}(x)\Big(\omega_{c2}+\omega_{0}+iv_{g}\frac{\partial} {\partial x}\Big)b_{L}^{\prime}(x)\\
&+b_{R}^{\prime\dagger}(x)\Big(\omega_{c2}+\omega_{0}-iv_{g}\frac{\partial} {\partial x}\Big)b_{R}^{\prime}(x)\Big\rbrack\\
&+\Big\lbrace\int_{-\infty}^{+\infty}dx\xi_{a}\left[a_{R}(x)+a_{L}(x)\right]|r_{1}r_{2} \rangle\langle g_{1}g_{2}|\\
&\times\big[\delta(x)+\delta(x-d)e^{i\theta_{1}}\big]+\text{H.c.}\Big\rbrace\\
&+\Big\lbrace\int_{-\infty}^{+\infty}dx\xi_{b}\left[b_{R}(x)+b_{L}(x)\right]|r_{1}r_{2} \rangle\langle g_{1}g_{2}|\\
&\times\big[\delta(x)+\delta(x-d)e^{i\theta_{2}}\big]+\text{H.c.}\Big\rbrace.
\label{HC_x^eff}
\end{split}
\end{equation}
Here, the high-frequency oscillation terms $e^{\pm i(\Delta_{c1}+\delta_{kb})t}$ and $e^{\pm i(\Delta_{c2}+\delta_{ka})t}$ have been neglected by assuming $|\Delta_{c1}+\delta_{kb}|\gg g_{b}\Omega_{c1}/\delta_{kb}$ and $|\Delta_{c2}+\delta_{ka}|\gg g_{a}\Omega_{c2}/\delta_{ka}$, while taking $\Delta_{c1}+\delta_{ka}\simeq0$ and $\Delta_{c2}+\delta_{kb}\simeq0$. This regime corresponds to the case in which detunings $\Delta_{c1}$ and $\Delta_{c2}$ of the two driving fields are sufficiently large in magnitude and opposite to each other. In this way, as shown in Fig.~\figpanel{fig6}{c}, the equivalent two-level system as a combination of the model in Fig.~\figpanel{fig1}{c} and the model in Fig.~\figpanel{fig4}{c} behaves like a giant atom again because it is coupled to both waveguide modes at two points, allowing for the occurence of self-interference effect for each waveguide mode.

With $H_{C_{x}}^{\text{eff}}$ in Eq.~(\ref{HC_x^eff}) and taking $|\tilde{\Psi}_{C_{x}}\rangle=|\tilde{\Psi}_{B_{x}}\rangle$ in Eq.~(\ref{Phi_B^eff}) as the corresponding eigenstate, for a single photon of waveguide mode $a$ incident from port 1 on the left, one can compute effective reflectivity, transmissivity, and backward and forward conversion efficiencies (see Supplementary III) in order as given below,
\begin{widetext}
\begin{equation}
\begin{split}
P_{1\rightarrow1}^{\text{eff}}=|p_{1\rightarrow1}^{\text{eff}}|^2&=\left|\frac{-i\Upsilon_{a}(1+e^{i\phi_{a}}e^{i\theta_{1}})(1+e^{i\phi_{a}}e^{-i\theta_{1}})}{\delta_{ka}+\Delta_{c1}+2i\gamma+2i\Upsilon_{a}(1+e^{i\phi_{a}}\text{cos}\theta_{1})+2i\Upsilon_{b}(1+e^{i\phi_{b}}\text{cos}\theta_{2})}\right|^{2},\\
P_{1\rightarrow2}^{\text{eff}}=|p_{1\rightarrow2}^{\text{eff}}|^2&=\left|\frac{\delta_{ka}+\Delta_{c1}+2i\gamma-2\Upsilon_{a}e^{i\theta_{1}}\text{sin}\phi_{a}+2i\Upsilon_{b}(1+e^{i\phi_{b}}\text{cos}\theta_{2})}{\delta_{ka}+\Delta_{c1}+2i\gamma+2i\Upsilon_{a}(1+e^{i\phi_{a}}\text{cos}\theta_{1})+2i\Upsilon_{b}(1+e^{i\phi_{b}}\text{cos}\theta_{2})}\right|^{2},\\
P_{1\rightarrow3}^{\text{eff}}=|p_{1\rightarrow3}^{\text{eff}}|^2&=\left|\frac{-i\sqrt{\Upsilon_{a}\Upsilon_{b}}(1+e^{i\phi_{b}}e^{i\theta_{2}})(1+e^{i\phi_{a}}e^{-i\theta_{1}})}{\delta_{ka}+\Delta_{c1}+2i\gamma+2i\Upsilon_{a}(1+e^{i\phi_{a}}\text{cos}\theta_{1})+2i\Upsilon_{b}(1+e^{i\phi_{b}}\text{cos}\theta_{2})}\right|^{2},\\
P_{1\rightarrow4}^{\text{eff}}=|p_{1\rightarrow4}^{\text{eff}}|^2&=\left|\frac{-i\sqrt{\Upsilon_{a}\Upsilon_{b}}(1+e^{-i\phi_{b}}e^{i\theta_{2}})(1+e^{i\phi_{a}}e^{-i\theta_{1}})}{\delta_{ka}+\Delta_{c1}+2i\gamma+2i\Upsilon_{a}(1+e^{i\phi_{a}}\text{cos}\theta_{1})+2i\Upsilon_{b}(1+e^{i\phi_{b}}\text{cos}\theta_{2})}\right|^{2}.
\label{P^eff}
\end{split}
\end{equation}
\end{widetext}
In these expressions, phase difference $\theta_{1}$ ($\theta_{2}$) between two driving points of the coherent field $\Omega_{c1}$ ($\Omega_{c2}$) will bring about chiral coupling effects together with phase difference $\phi_{a}$ ($\phi_{b}$) between two coupling points of the waveguide mode $a$ ($b$). Then, the scattered photon can be routed toward a selected port while suppressing the probabilities toward other ports, which definitely improves reflectivity $P_{1\rightarrow1}^{\text{eff}}$, transmissivity $P_{1\rightarrow2}^{\text{eff}}$, or frequency conversion efficiencies $P_{1\rightarrow3}^{\text{eff}}$ and $P_{1\rightarrow4}^{\text{eff}}$ on demand.

\begin{figure}[pth]
\centering
\includegraphics[width=8.4 cm]{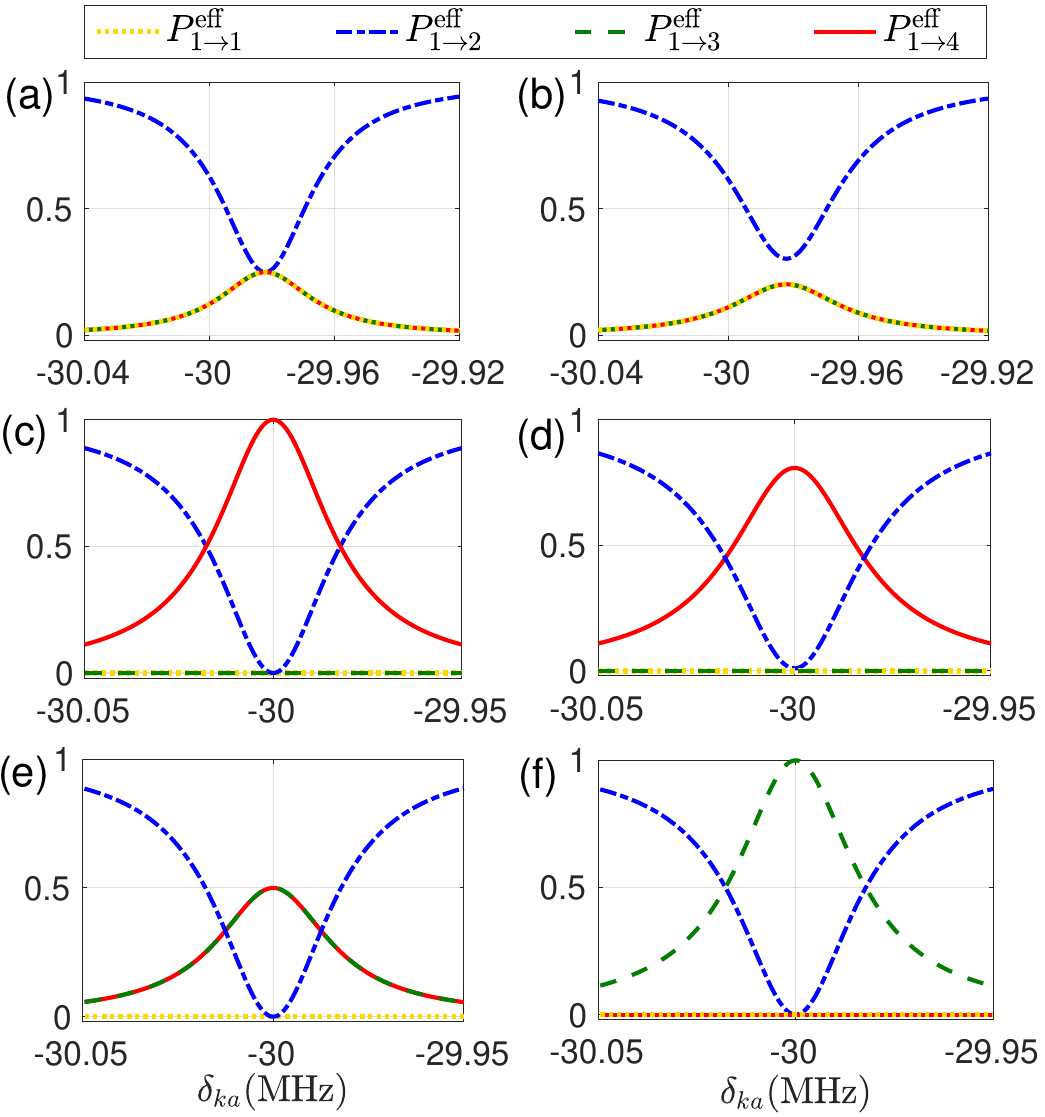}
\caption{Effective reflectivity $P^{\text{eff}}_{1\rightarrow1}$, transmissivity $P^{\text{eff}}_{1\rightarrow2}$, and backward $P^{\text{eff}}_{1\rightarrow3}$ and forward $P^{\text{eff}}_{1\rightarrow4}$ conversion efficiencies versus detuning $\delta_{ka}$ with (a) $\gamma=0$, $\phi_{b}=\pi/2$, and $\theta_{1}=\theta_{2}=0$; (b) $\gamma=1$ kHz, $\phi_{b}=\pi/2$, and $\theta_{1}=\theta_{2}=0$; (c) $\gamma=0$, $\phi_{b}=\pi/2$, and $\theta_{1}=\theta_{2}=\pi/2$; (d) $\gamma=1$ kHz, $\phi_{b}=\pi/2$, and $\theta_{1}=\theta_{2}=\pi/2$; (e) $\gamma=0$, $\phi_{b}=\pi$, and $\theta_{1}=\theta_{2}=\pi/2$; (f) $\gamma=0$, $\phi_{b}=3\pi/2$, and $\theta_{1}=\theta_{2}=\pi/2$. Other parameters are $\phi_{a}=\pi/2$, $\Gamma_{a}=\Gamma_{b}=1$ MHz, $\Omega_{c1}=\Omega_{c2}=2$ MHz, and $\Delta_{c1}=-\Delta_{c2}=30$ MHz, and $V_{6}=20$ GHz.}
\label{fig7}
\end{figure}

Figure~\ref{fig7} shows typical spectra of the effective reflectivity, transmissivity, and backward and forward conversion efficiencies for different values of relevant phase differences and intrinsic decay rates. It can be seen from Figs.~\figpanel{fig7}{a} and~\figpanel{fig7}{b} that, when the two coherent fields are at normal incidence (i.e., $\theta_{1}=\theta_{2}=0$), the equivalent couplings are nonchiral and the optimal frequency conversion efficiency $P^{\rm{eff}}_{1\rightarrow3}+P^{\rm{eff}}_{1\rightarrow4}=0.5$ is the same as that found in the last section.
For the perfectly chiral case corresponding to $\theta_{1}=\theta_{2}=\pi/2$ and $\phi_{a}=\phi_{b}=\pi/2$ shown in Figs.~\figpanel{fig7}{c} and~\figpanel{fig7}{d}, the forward conversion efficiency can approach unity around $\delta_{ka}\simeq-\Delta_{c1}$ with $\gamma=0$, though in general being smaller than unity for $\gamma\ne0$. This is because both reflectivity $P^{\rm{eff}}_{1\rightarrow1}$ and backward conversion efficiency $P^{\rm{eff}}_{1\rightarrow3}=0$ remains zero, independent of $\delta_{ka}$, as a result of perfect destructive interference contributed by $1+e^{i\phi_{a}}e^{i\theta_{1}}$ and $1+e^{i\phi_{b}}e^{i\theta_{2}}$, respectively, as can be seen from Eq.~(\ref{P^eff}). Hence, our waveguide-coupled Rydberg-atom system is an excellent platform to realize high-efficiency frequency conversion, needing neither multiple-level atomic configurations nor specific devices like Sagnac interferometers \cite{sag}. It is also important that one can adjust the output ports of mode $b$ converted from mode $a$ by tuning $\phi_{b}$ to attain $P^{\rm{eff}}_{1\rightarrow3}=P^{\rm{eff}}_{1\rightarrow4}=0.5$ with $\phi_{b}=\pi$ shown in Fig.~\figpanel{fig7}{e}, or $P^{\rm{eff}}_{1\rightarrow3}=1$ with $\phi_{b}=3\pi/2$ shown in Fig.~\figpanel{fig7}{f}, in the ideal case with $\gamma=0$. As a corollary of this observation, the equivalent chirality of waveguide modes $a$ and $b$ can be controlled separately with different values of $\phi_{a}$ and $\phi_{b}$. Indeed, the functions of $\phi_{a}$ and $\theta_{1}$ ($\phi_{b}$ and $\theta_{2}$) are similar, both of which can be used to tune the chirality of waveguide mode $a$ ($b$) as depicted in Fig.~\figpanel{fig3}{d} only for waveguide mode $a$.

\begin{figure}[ptb]
\centering
\includegraphics[width=8.5 cm]{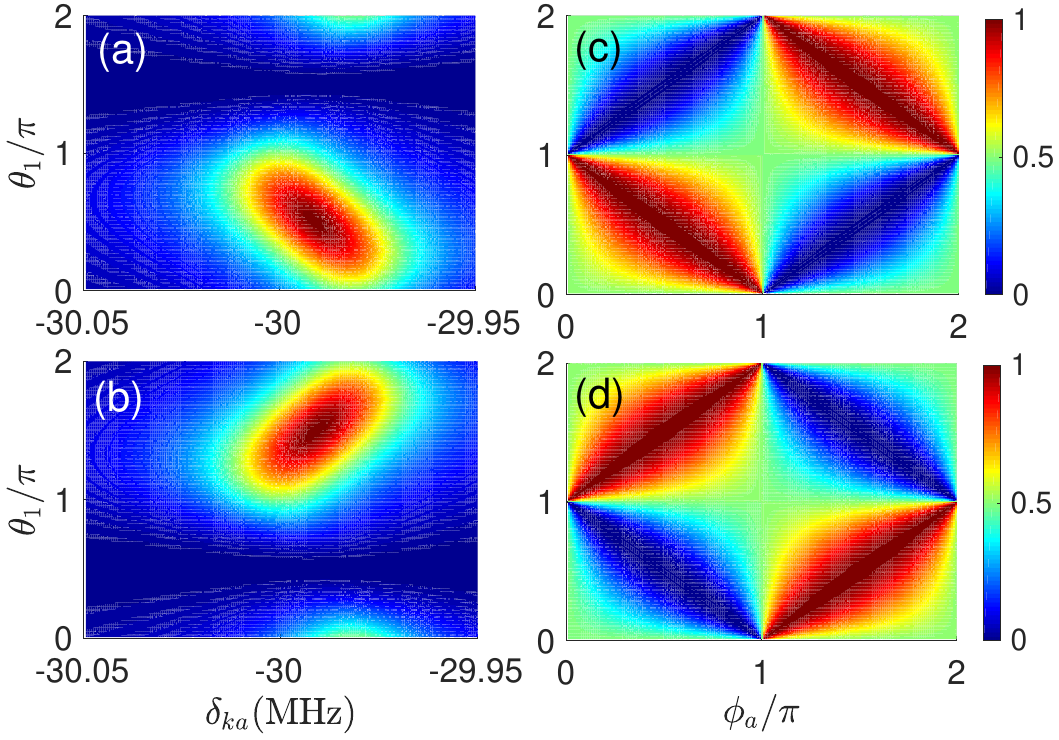}
\caption{Effective total conversion efficiency (a, c) $P^{\text{eff}}_{1\rightarrow3}+P^{\text{eff}}_{1\rightarrow4}$ and (b, d) $P^{\text{eff}}_{2\rightarrow3}+P^{\text{eff}}_{2\rightarrow4}$ versus detuning $\delta_{ka}$ and phase $\theta_{1}$ with $\phi_{a}=\phi_{b}=\pi/2$ (a, b); versus phases $\phi_{b}=\phi_{a}$ and $\theta_{1}$ with $\delta_{ka}=-30$ MHz (c, d). Other parameters are the same as in Fig.~\ref{fig7} except $\theta_{2}=0$ and $\gamma=0$.}
\label{fig8}
\end{figure}

Based on this chiral coupling mechanism, it is natural to consider the possible realization of asymmetric frequency conversion by comparing scattering behaviors of a left-incident photon with that of a right-incident photon in the same waveguide mode. In fact, for a photon incident from port 2 in mode $a$, the analytical expression of total conversion efficiency $P^{\text{eff}}_{2\rightarrow3}+P^{\text{eff}}_{2\rightarrow4}$ will be identical to that of $P^{\text{eff}}_{1\rightarrow3}+P^{\text{eff}}_{1\rightarrow4}$ if we make the replacement $\theta_{1,2}\rightarrow -\theta_{1,2}$. It can be found from Figs.~\figpanel{fig8}{a} and~\figpanel{fig8}{b}, with $\phi_{a}=\phi_{b}=\pi/2$ and $\theta_{2}=0$, the spectra of above two conversion efficiencies into mode $b$ have an inverse dependence on $\theta_{1}$ for the left-incident and right-incident photons in mode $a$. That means, the chirality of mode $a$ is enough to realize asymmetric frequency conversion since it can fully determine whether a photon in mode $a$ can be first absorbed by the two atoms and then converted into mode $b$ while the chirality of mode $b$ will only determine the output port (3 or 4) of the photon converted into mode $b$. Note also that $P^{\text{eff}}_{2\rightarrow3}=P^{\text{eff}}_{2\rightarrow4}$ always holds in Figs.~\figpanel{fig8}{a} and~\figpanel{fig8}{b} due to the fact that $\theta_{2}=0$ refers to the nonchiral case. Figures~\figpanel{fig8}{c} and~\figpanel{fig8}{d} show that $\phi_{a}$ and $\phi_{b}$ will also determine the chirality, and with $\phi_{a}=\phi_{b}=n\pi$ the frequency conversion turns out to be symmetric, which is consistent with Fig.~\figpanel{fig3}{d}.

\section{Conclusions}
Waveguide-coupled Rydberg atoms represent a relatively new form of hybrid quantum systems, made of individual components with complementary characteristics, e.g., long coherence times, flexible all-optical control, available transition frequencies; these systems are of interest to a wide range of areas, from quantum computation and communication to quantum sensing. The present work represents a detailed and systematic account of the large degree of control over the range and the nature of equivalent chiral couplings and nonreciprocal scattering enabled by two Rydberg atoms coupled to a waveguide so as to form a giant atom. Our waveguide-coupled Rydberg giant-atom platform may also be adapted to work as a frequency converter with an efficiency that can exhibit a strong asymmetry and be significantly enhanced via chiral couplings. It's certainly noteworthy to mention here that our model can be easily expanded to accomplish a multi-frequency conversion. This entails incorporating additional driving fields with varying frequencies, coupled to the upper transitions, to successfully implement two-photon transitions. We believe that our proposal provides an innovative approach to chiral giant-atom physics and one that may lead to novel results on the Rydberg nonlinearity at optical frequencies.

\section*{ACKNOWLEDGMENTS}
This work is supported by the National Natural Science
Foundation of China (Nos.~12074061 and 62375047) and the Italian
PNRR MURR (No.~PE0000023-NQSTI). YTC and JHW would like to thank the
hospitality at Scuola Normale Superiore in Pisa.

\begin{widetext}

\section*{SUPPLEMENTAL MATERIAL}

\begin{adjustwidth}{1.5cm}{1.5cm}
\qquad
This supplementary material gives further calculation details on the first model shown in Fig.~1 of the main text with nonreciprocal transmissions (Sec. I), the second model shown in Fig.~4 of the main text with symmetric frequency conversion (Sec. II), and the third model shown in Fig.~6 of the main text with asymmetric frequency conversion (Sec. III) that are omitted in the main text.
\end{adjustwidth}

\section*{I.\quad nonreciprocal transmission with one single field}

\subsection{Two Rydberg atoms without any constraints}

Here we would like to show how to apply the Bethe-ansatz approach to solve the scattering problem for the two Rydberg atoms.
By solving the eigenequation $H_{A_{x}}|\Psi_{A_{x}}\rangle=\omega_{ka}|\Psi_{A_{x}}\rangle$ from Eq.~(2) and Eq.~(3) in the main text, one can obtain
\begin{equation}
\begin{split}
\omega_{ka}\Phi_{aR}^{A}(x)&=e^{ik_{0}x}\Big(\omega_{0}-iv_{g}\frac{\partial}{\partial x}\Big)\Phi_{aR}^{A}(x)e^{-ik_{0}x}+g_{a}\delta(x)u_{b}^{A}+g_{a}\delta(x-d)u_{c}^{A},\\
\omega_{ka}\Phi_{aL}^{A}(x)&=e^{-ik_{0}x}\Big(\omega_{0}+iv_{g}\frac{\partial}{\partial x}\Big)\Phi_{aL}^{A}(x)e^{ik_{0}x}+g_{a}\delta(x)u_{b}^{A}+g_{a}\delta(x-d)u_{c}^{A},\\
\omega_{ka}u_{b}^{A}&=\Big(\omega_{e}-i\gamma\Big)u_{b}^{A}+g_{a}\big\lbrack\Phi_{R}^{A}(0)+\Phi_{L}^{A}(0)\big\rbrack+\Omega_{c1}u_{d}^{A},\\
\omega_{ka}u_{c}^{A}&=\Big(\omega_{e}-i\gamma\Big)u_{c}^{A}+g_{a}\big\lbrack\Phi_{R}^{A}(d)+\Phi_{L}^{A}(d)\big\rbrack+\Omega_{c1}e^{-i\theta_{1}}u_{d}^{A},\\
\omega_{ka}u_{d}^{A}&=\Big(\omega_{e}-\Delta_{c1}-2i\gamma\Big)u_{d}^{A}+\Omega_{c1}u_{b}^{A}+\Omega_{c1}e^{i\theta_{1}}u_{c}^{A},
\end{split}
\tag{S1}
\label{A1}
\end{equation}
where $\omega_{ka}$ is the frequency of one incident photon satisfying $\omega_{ka}=\omega_{0}+(k_{a}-k_{0})v_{g}$.
For the left-incident case (from port 1), substituting the wave functions $\Phi_{aR,aL}^{A}(x)$ in Eq.~(4) into Eq.~(\ref{A1}), we obtain
\begin{equation}
\begin{split}
0&=-iv_{g}(A_{1}-1)+g_{a}u_{b}^{A},\\
0&=-iv_{g}(t_{1\rightarrow2}-A_{1})e^{i\phi_{a}}+g_{a}u_{c}^{A},\\
0&=-iv_{g}(r_{1\rightarrow1}-A_{2})+g_{a}u_{b}^{A},\\
0&=-iv_{g}A_{2}e^{-i\phi_{a}}+g_{a}u_{c}^{A},\\
0&=\frac{g_{a}}{2}(A_{1}+A_{2}+r_{1\rightarrow1}+1)+\Omega_{c1}u_{d}^{A}-(\delta_{ka}+i\gamma)u_{b}^{A},\\
0&=\frac{g_{a}}{2}(A_{1}e^{i\phi_{a}}+A_{2}e^{-i\phi_{a}}+t_{1\rightarrow2}e^{i\phi_{a}})+\Omega_{c1}e^{-i\theta_{1}}u_{d}^{A}-(\delta_{ka}+i\gamma)u_{c}^{A},\\
0&=\Omega_{c1}u_{b}^{A}+\Omega_{c1}e^{i\theta_{1}}u_{c}^{A}-(\delta_{ka}+\Delta_{c1}+2i\gamma)u_{d}^{A}
\label{A3}
\end{split}
\tag{S2}
\end{equation}
with $\phi_{a}=k_{a}d=k_{0}d+(\omega_{ka}-\omega_{0})d/v_{g}$ that can be seen as a constant in the Markovian regime \cite{scp2022} and $\delta_{ka}=\omega_{ka}-\omega_{e}$.
Then solving Eq.~(\ref{A3}) numerically, one can get the transmissivity from port 1 to port 2 as $T_{1\rightarrow 2}=|t_{1\rightarrow 2}|^{2}$.
If the photon is incident from the right port, i.e., port 2, then one can similarly substitute Eq.~(5) into Eq.~(\ref{A1}) to obtain the transmissivity from port 2 to port 1. It can be found that 
all the results are the same compared with the left-incident case yet with an opposite sign of the phase difference, i.e., $\theta_{1}\leftrightarrow-\theta_{1}$.

\subsection{Equivalent giant atom with large detunings}

Based on the effective Hamiltonian theory \cite{selimination1,selimination2,seff1}, for the model shown in Fig.~1 in the main text, we first need to obtain the Hamiltonian in the interaction picture from $H_{A_{k}}$ in Eq.~(1) as
\begin{equation}
\begin{split}
\mathcal{H}_{A_{k}}(t)&=i\frac{dU_{1}^{\dagger}}{dt}U_{1}+U_{1}^{\dagger}H_{A_{k}}U_{1}\\
&=\int dk_{a}g_{a}a_{k}(|r_{1}g_{2} \rangle\langle g_{1}g_{2}|+|g_{1}r_{2} \rangle\langle g_{1}g_{2}|e^{ik_{a}d})e^{-i\delta_{ka}t}+\Omega_{c1}
e^{-i\Delta_{c1}t}(|r_{1}r_{2} \rangle\langle r_{1}g_{2}|+e^{i\theta_{1}}|r_{1}r_{2} \rangle\langle g_{1}r_{2}|)+\text{H.c.},
\end{split}
\tag{S3}
\label{A4}
\end{equation}
where the unitary operator $U_{1}=e^{-iH_{A_{1}}t}$ with $H_{A_{1}}=\omega_{e}(|g_{1}r_{2}\rangle\langle g_{1}r_{2}|+|r_{1}g_{2}\rangle\langle r_{1}g_{2}|)+(2\omega_{e}+V_{6})|r_{1}r_{2}\rangle\langle r_{1}r_{2}|
+\int dk_{a}\omega_{ka}a_{k}^{\dagger}a_{k}$.
Then the effective Hamiltonian in the interaction picture can be obtained as
\begin{equation}
\begin{split}
\mathcal{H}_{A_{k}}^{\rm{eff}}(t)&=-i\mathcal{H}_{A_{k}}(t)\int_{0}^{t}\mathcal{H}_{A_{k}}(t')dt'\\
&\simeq\frac{2g_{a}^{2}}{\delta_{ka}}\int dk_{a}a_{k}a_{k}^{\dagger}|g_{1}g_{2}\rangle \langle g_{1}g_{2}|-\frac{2\Omega_{c1}^{2}}{\Delta_{c1}}|r_{1}r_{2}\rangle \langle r_{1}r_{2}|-2i\gamma|r_{1}r_{2}\rangle \langle r_{1}r_{2}|\\
&\quad\,+\frac{g_{a}\Omega_{c1}e^{i\theta_{1}}}{\delta_{ka}}\int dk_{a}a_{k}e^{-i(\delta_{ka}+\Delta_{c1})t}|r_{1}r_{2}\rangle
\langle g_{1}g_{2}|(1+e^{ik_{a}d})\\
&\quad\,-\frac{g\Omega_{c1}e^{-i\theta_{1}}}{\Delta_{c1}}\int dk_{a}a_{k}^{\dagger}e^{i(\delta_{ka}+\Delta_{c1})t}|g_{1}g_{2}\rangle
\langle r_{1}r_{2}|(1+e^{-ik_{a}d})+\cdots,
\end{split}
\tag{S4}
\label{A5}
\end{equation}
where we have omitted a few terms related to the single-excitation states $|r_{1}g_{2}\rangle$ and $|g_{1}r_{2}\rangle$ since they are
decoupled from other states and only interact with each other.
Assuming $\Delta_{c1}+\delta_{ka}\simeq0$ and
$|\Delta_{c1}|, |\delta_{ka}|\gg\Omega_{c1},g_{a}$, we will have
 the effective Hamiltonian in the Schr\"{o}dinger picture as 
\begin{equation}
\begin{split}
H_{A_{k}}^{\text{eff}}&=i\frac{dU_{2}^{\dagger}}{dt}U_{2}+U_{2}^{\dagger}\mathcal{H}_{A_{k}}^{\rm{eff}}(t)U_{2}\\
&=(2\omega_{e}+V_{6}-\frac{2\Omega_{c1}^{2}}{\Delta_{c1}}-2i\gamma)|r_{1}r_{2}\rangle \langle r_{1}r_{2}|+\int dk_{a}(\omega_{ka}+\omega_{c1})a_{k}^{\dagger}a_{k}+\int dk_{a}\xi_{a}a_{k}(1+e^{ik_{a}d}e^{i\theta_{1}})|r_{1}r_{2}\rangle
\langle g_{1}g_{2}|+\text{H.c.},
\label{A6}
\end{split}
\tag{S5}
\end{equation}
where $\xi_{a}=-g_{a}\Omega_{c1}/\Delta_{c1}$ and $U_{2}=e^{-iH_{A_{2}}t}$ with $H_{A_{2}}=-(2\omega_{e}+V_{6})|r_{1}r_{2}\rangle\langle r_{1}r_{2}|
-\int dk_{a}(\omega_{ka}+\omega_{c1})a_{k}^{\dagger}a_{k}$. 
Then this effective momentum-space Hamiltonian can be transformed to the real space as in Eq.~(6) in the main text. 

Then, with the same procedure, by solving the eigenequation $H_{A_{x}}^{\text{eff}}|\tilde{\Psi}_{A_{x}}\rangle=(\omega_{ka}+\omega_{c1})|\tilde{\Psi}_{A_{x}}\rangle$ from Eq.~(6) and Eq.~(7) in the main text, one can get the following equations
\begin{equation}
\begin{split}
(\omega_{ka}+\omega_{c1})\tilde{\Phi}^{A}_{aR}(x)&=e^{ik_{0}x}\Big(\omega_{c1}+\omega_{0}-iv_{g}\frac{\partial}{\partial x}\Big)\tilde{\Phi}^{A}_{aR}(x)e^{-ik_{0}x}+\xi_{a}\Big\lbrack\delta(x)+e^{i\theta_{1}}\delta(x-d)\Big\rbrack \tilde{u}_{d}^{A},\\
(\omega_{ka}+\omega_{c1})\tilde{\Phi}^{A}_{aL}(x)&=e^{-ik_{0}x}\Big(\omega_{c1}+\omega_{0}+iv_{g}\frac{\partial}{\partial x}\Big)\tilde{\Phi}^{A}_{aL}(x)e^{ik_{0}x}+\xi_{a}\Big\lbrack\delta(x)+e^{i\theta_{1}}\delta(x-d)\Big\rbrack \tilde{u}_{d}^{A},\\
(\omega_{ka}+\omega_{c1}))\tilde{u}^{A}_{d}&=(2\omega_{e}+V_{6}-\frac{2\Omega_{c1}^{2}}{\Delta_{c1}}-2i\gamma)\tilde{u}^{A}_{d}+\xi_{a}\big\lbrack\tilde{\Phi}^{A}_{aR}(0)+\tilde{\Phi}^{A}_{aL}(0)\big\rbrack+\xi_{a}e^{-i\theta_{1}}\big\lbrack\tilde{\Phi}^{A}_{aR}(d)+\tilde{\Phi}^{A}_{aL}(d)\big\rbrack
\label{A8}
\end{split}
\tag{S6}
\end{equation}
and
\begin{equation}
\begin{split}
0&=-iv_{g}(A_{1}^{\rm{eff}}-1)+\xi_{a}\tilde{u}^{A}_{d},\\
0&=-iv_{g}(t^{\rm{eff}}_{1\rightarrow2}-A_{1}^{\rm{eff}})e^{i\phi_{a}}+\xi_{a}e^{i\theta_{1}}\tilde{u}^{A}_{d},\\
0&=-iv_{g}(r^{\rm{eff}}_{1\rightarrow1}-A_{2}^{\rm{eff}})+\xi_{a}\tilde{u}^{A}_{d},\\
0&=-iv_{g}A_{2}^{\rm{eff}}e^{-i\phi_{a}}+\xi_{a}e^{i\theta_{1}}\tilde{u}^{A}_{d},\\
0&=\frac{\xi_{a}}{2}(A_{1}^{\rm{eff}}+A_{2}^{\rm{eff}}+r^{\rm{eff}}_{1\rightarrow1}+1)+\frac{\xi_{a}}{2}e^{-i\theta_{1}}(A_{1}^{\rm{eff}}e^{i\phi_{a}}+A_{2}^{\rm{eff}}e^{-i\phi_{a}}+t^{\rm{eff}}_{1\rightarrow2}e^{i\phi_{a}})-(\delta_{ka}+\Delta_{c1}+\frac{2\Omega_{c1}^{2}}{\Delta_{c1}}+2i\gamma)\tilde{u}^{A}_{d}.
\label{A9}
\end{split}
\tag{S7}
\end{equation}
The effective transmissivity $T_{1\rightarrow 2}^{\text{eff}}=|t_{1\rightarrow 2}^{\text{eff}}|^{2}$ of Eq.~(8) can be obtained from solving Eq.~(\ref{A9}). The transmissivity $T_{2\rightarrow 1}^{\text{eff}}=|t_{2\rightarrow 1}^{\text{eff}}|^{2}$ in Eq.~(8) of a right-incident photon is dealt with in the same way as above.

\subsection{Continuous-coupling case based on dynamical solution method}

In the case of continuous coupling, elucidating the wave function ansatz in real space can be challenging, particularly due to the unclear delineation of the coupling region.
Therefore, we are considering the application of the dynamical solution method based on the momentum-space description~\cite{sliao2016,sdecay1}. Additionally, we illustrate the process of obtaining transmissivities for the point-like coupling case once again, providing an example.
Based on the effective Hamiltonian in Eq.~(\ref{A6}),
the time-evolving state at time $t$ in the single-excitation invariant subspace can be written as:
\begin{equation}
\begin{split}
|\Psi_{A_{k}}(t)\rangle&=\int dk_{a}\big\lbrack c_{1}(t)a_{kR}^{\dagger}+c_{2}(t)a_{kL}^{\dagger}\big\rbrack|0,g_{1}g_{2}\rangle+\tilde{u}^{A}_{d}(t)|0,r_{1}r_{2}\rangle.
\label{D1}
\end{split}
\tag{S8}
\end{equation}
Therefore, we have the dynamical equations for the equivalent giant atom
 \begin{equation}
\begin{split}
\partial_{t}\tilde{u}^{A}_{d}(t)&=-(i\omega_{e}-i\Delta_{c1}-i\frac{2\Omega_{c1}^{2}}{\Delta_{c1}}+2\gamma)\tilde{u}^{A}_{d}(t)-i\xi\int dk_{a}\big(1+e^{ik_{a}d}e^{-i\theta_{1}}\big)c_{1}(t)-i\xi\int dk_{a}\big(1+e^{-ik_{a}d}e^{-i\theta_{1}}\big)c_{2}(t),\\
\partial_{t}c_{1}(t)&=-i\omega_{ka}c_{1}(t)-i\tilde{u}^{A}_{d}(t)\xi\big(1+e^{-ik_{a}d}e^{i\theta_{1}}\big),\\
\partial_{t}c_{2}(t)&=-i\omega_{ka}c_{2}(t)-i\tilde{u}^{A}_{d}(t)\xi\big(1+e^{ik_{a}d}e^{i\theta_{1}}\big)\\
\label{D2}
\end{split}
\tag{S9}
\end{equation}
with $\xi=\xi_{a}/\sqrt{2\pi}$. $c_{1}(t)$ and $c_{2}(t)$ refer to the right- and left-propagating fields, respectively.
Integrating the latter two equations, we have
 \begin{equation}
\begin{split}
c_{1}(t)&=e^{-i\omega_{ka}t}\left[c_{1}(0)-i\xi(1+e^{-ik_{a}d}e^{i\theta_{1}})\int_{0}^{t}dt'\tilde{u}^{A}_{d}(t')e^{i\omega_{ka}t'}\right],\\
c_{2}(t)&=e^{-i\omega_{ka}t}\left[c_{2}(0)-i\xi(1+e^{ik_{a}d}e^{i\theta_{1}})\int_{0}^{t}dt'\tilde{u}^{A}_{d}(t')e^{i\omega_{ka}t'}\right].
\label{D3}
\end{split}
\tag{S10}
\end{equation}
Inserting these two equations into the dynamical equation of $\tilde{u}^{A}_{d}(t)$ in Eq.~(\ref{D2}) and after arranging, we have 
\begin{equation}
\begin{split}
\partial_{t}\tilde{u}^{A}_{d}(t)
&=-(i\omega_{e}-i\Delta_{c1}-i\frac{2\Omega_{c1}^{2}}{\Delta_{c1}}+2\gamma)\tilde{u}^{A}_{d}(t)-i\frac{\xi}{v_{g}}\int d\omega_{ka}\big(1+e^{i\omega_{ka}d/v_{g}}e^{-i\theta_{1}}\big)e^{-i\omega_{ka}t}c_{1}(0)\\
&\,\quad-i\frac{\xi}{v_{g}}\int d\omega_{ka}\big(1+e^{-i\omega_{ka}d/v_{g}}e^{-i\theta_{1}}\big)e^{-i\omega_{ka}t}c_{2}(0)\\
&\,\quad-\frac{\xi^{2}}{v_{g}}\int_{0}^{t}dt'\tilde{u}^{A}_{d}(t')\int d\omega_{ka}\Big[2e^{-i\omega_{ka}(t-t')}+e^{-i\omega_{ka}(t-t'-d/v_{g})}e^{-i\theta_{1}}+e^{-i\omega_{ka}(t-t'+d/v_{g})}e^{i\theta_{1}}\Big]\\
&\,\quad-\frac{\xi^{2}}{v_{g}}\int_{0}^{t}dt'\tilde{u}^{A}_{d}(t')\int d\omega_{ka}\Big[2e^{-i\omega_{ka}(t-t')}+e^{-i\omega_{ka}(t-t'-d/v_{g})}e^{i\theta_{1}}+e^{-i\omega_{ka}(t-t'+d/v_{g})}e^{-i\theta_{1}}\Big]\\
&=-(i\omega_{e}-i\Delta_{c1}-i\frac{2\Omega_{c1}^{2}}{\Delta_{c1}}+2\gamma)\tilde{u}^{A}_{d}(t)-i\frac{\xi}{v_{g}}\int d\omega_{ka}\big(1+e^{i\omega_{ka}d/v_{g}}e^{-i\theta_{1}}\big)e^{-i\omega_{ka}t}c_{1}(0)\\
&\,\quad-i\frac{\xi}{v_{g}}\int d\omega_{ka}\big(1+e^{-i\omega_{ka}d/v_{g}}e^{-i\theta_{1}}\big)e^{-i\omega_{ka}t}c_{2}(0)\\
&\,\quad-\frac{2\pi \xi^{2}}{v_{g}}\int_{0}^{t}dt'\tilde{u}^{A}_{d}(t')\Big[2\delta(t-t')+\delta(t'-t+d/v_{g}) e^{-i\theta_{1}}+\delta(t'-t-d/v_{g}) e^{i\theta_{1}}\Big]\\
&\,\quad-\frac{2\pi \xi^{2}}{v_{g}}\int_{0}^{t}dt'\tilde{u}^{A}_{d}(t')\Big[2\delta(t-t')+\delta(t'-t+d/v_{g}) e^{i\theta_{1}}+\delta(t'-t-d/v_{g}) e^{-i\theta_{1}}\Big]\\
&=-(i\omega_{e}-i\Delta_{c1}-i\frac{2\Omega_{c1}^{2}}{\Delta_{c1}}+2\gamma)\tilde{u}^{A}_{d}(t)-i\frac{\xi}{v_{g}}\int d\omega_{ka}\big(1+e^{i\omega_{ka}d/v_{g}}e^{-i\theta_{1}}\big)e^{-i\omega_{ka}t}c_{1}(0)\\
&\,\,\quad-i\frac{\xi}{v_{g}}\int d\omega_{ka}\big(1+e^{-i\omega_{ka}d/v_{g}}e^{-i\theta_{1}}\big)e^{-i\omega_{ka}t}c_{2}(0)\\
&\,\,\quad-\frac{4\pi \xi^{2}}{v_{g}}\tilde{u}^{A}_{d}(t)-\frac{2\pi \xi^{2}}{v_{g}}\Theta(t-d/v_{g})\tilde{u}^{A}_{d}(t-d/v_{g}) e^{-i\theta_{1}}-\frac{2\pi \xi^{2}}{v_{g}}\Theta(t-d/v_{g})\tilde{u}^{A}_{d}(t-d/v_{g}) e^{i\theta_{1}}.
\label{D4}
\end{split}
\tag{S11}
\end{equation}
Here, we have assumed $\tilde{u}^{A}_{d}(t)=0$ for $t<0$ and thus neglected the Heaviside step function $\Theta(t-d/v_{g})$.
By Fourier-transforming this equation and with the initial condition $\tilde{u}^{A}_{d}(0)=0$,
we have 
 \begin{equation}
\begin{split}
-i\tilde{u}^{A}_{d}(\omega_{ka})&=-(i\omega_{e}-i\Delta_{c1}-i\frac{2\Omega_{c1}^{2}}{\Delta_{c1}}+2\gamma)\tilde{u}^{A}_{d}(\omega_{ka})-i\frac{2\pi \xi}{v_{g}}\big(1+e^{i\omega_{ka}d/v_{g}}e^{-i\theta_{1}}\big)c_{1}(0)-i\frac{2\pi \xi}{v_{g}}\big(1+e^{-i\omega_{ka}d/v_{g}}e^{-i\theta_{1}}\big)c_{2}(0)\\
&\,\quad-\frac{4\pi \xi^{2}}{v_{g}}\tilde{u}^{A}_{d}(\omega_{ka})-\frac{4\pi \xi^{2}}{v_{g}}\tilde{u}^{A}_{d}(\omega_{ka})e^{i\omega_{a}d/v_{g}} \text{cos}\theta_{1},\\
\label{D5}
\end{split}
\tag{S12}
\end{equation}
and
\begin{equation}
\begin{split}
\tilde{u}^{A}_{d}(t)&=\int d\omega_{ka}\frac{\frac{\xi}{v_{g}}\big(1+e^{i\omega_{ka}d/v_{g}} e^{-i\theta_{1}}\big)c_{1}(0)+\frac{\xi}{v_{g}}\big(1+e^{-i\omega_{ka}d/v_{g}} e^{-i\theta_{1}}\big)c_{2}(0)}{\omega_{ka}-\omega_{e}+\Delta_{c1}+\frac{2\Omega_{c1}^{2}}{\Delta_{c1}}+2i\gamma+i\frac{4\pi \xi^{2}}{v_{g}}(1+e^{i\omega_{ka}d/v_{g}} \text{cos}\theta_{1})}e^{-i\omega_{ka}t}.
\label{D6}
\end{split}
\tag{S13}
\end{equation}
For the left-incident photon with $c_{1}(0)\neq0$ and $c_{2}(0)=0$, we substitute this equation into the equation of $c_{1}(t)$
\begin{equation}
\begin{split}
c_{1}(t)&=e^{-i\omega_{ka}t}c_{1}(0)-i\frac{\xi^{2}}{v_{g}}\big(1+e^{-i\omega_{ka}d/v_{g}} e^{i\theta_{1}}\big)\int d\omega'_{ka}\frac{\big(1+e^{i\omega'_{ka}d/v_{g}} e^{-i\theta_{1}}\big)c_{1}(0)}{\omega'_{ka}-\omega_{e}+\Delta_{c1}+\frac{2\Omega_{c1}^{2}}{\Delta_{c1}}+2i\gamma+i\frac{4\pi g^{2}}{v_{g}}(1+e^{i\omega'_{ka}d/v_{g}} \text{cos}\theta_{1}})\\
&\,\quad\times e^{-i\omega_{ka}t}\int_{0}^{t}dt'\tilde{u}^{A}_{d}(t')e^{i(\omega_{ka}-\omega'_{ka})t'},\\
c_{1}(\infty)&=e^{-i\omega_{ka}t}c_{1}(0)\left(1-i\frac{4\pi \xi^{2}}{v_{g}}\frac{1+\text{cos}(\omega_{ka}d/v_{g}) \text{cos}\theta_{1}+\text{sin}(\omega_{ka}d/v_{g})\text{sin}\theta_{1}}{\omega_{ka}-\omega_{e}+\Delta_{c1}+\frac{2\Omega_{c1}^{2}}{\Delta_{c1}}+2i\gamma+i\frac{4\pi \xi^{2}}{v_{g}}(1+e^{i\omega_{ka}d/v_{g}} \text{cos}\theta_{1})}\right),
\label{D7}
\end{split}
\tag{S14}
\end{equation}
which is in the long-time limit.
Therefore, the transmissivity $T_{1\rightarrow2}^{\text{eff}}\equiv\frac{|c_{1}(\infty)|^{2}}{|c_{1}(0)|^{2}}$ can be expressed as
\begin{equation}
\begin{split}
T_{1\rightarrow2}^{\text{eff}}&=\left|\frac{\delta_{ka}+\Delta_{c1}+\frac{2\Omega_{c1}^{2}}{\Delta_{c1}}+2i\gamma-2\Upsilon_{a}e^{i\theta_{1}}\text{sin}\phi_{a}}
{\delta_{ka}+\Delta_{c1}+\frac{2\Omega_{c1}^{2}}{\Delta_{c1}}+2i\gamma+2i\Upsilon_{a}(1+ e^{i\phi_{a}}\text{cos}\theta_{1})}\right|^{2}
\label{D8}
\end{split}
\tag{S15}
\end{equation}
with $\Upsilon_{a}=\xi_{a}^{2}/v_{g}$, which is the same as that in Eq.~(8) in the main text calculated from the Bethe-ansantz method.

Taking continuous coupling into account, consider the example with the coupling distribution represented by exponential functions as $\nu_{1}(x)=\frac{\xi}{\Lambda}e^{-\frac{2}{\Theta}|x|}$ at $x=0$ and $\nu_{2}(x)=\frac{\xi}{\Lambda}e^{-\frac{2}{\Theta}|x-d|}$ at $x=d$, the momentum-space Hamiltonian can be written as 
\begin{equation}
\begin{split}
H_{A_{k}}^{\text{eff}_{c}}&=(2\omega_{e}+V_{6}-\frac{2\Omega_{c1}^{2}}{\Delta_{c1}}-2i\gamma)|r_{1}r_{2} \rangle\langle r_{1}r_{2}|+\int dk_{a}(\omega_{ka}+\omega_{c1})(a_{kL}^{\dagger}a_{kL}+a_{kR}^{\dagger}a_{kR})\\
&\,\quad+\int dx\int dk_{a}\Big[a_{kL}^{\dagger}|g\rangle\langle e|\big[\nu_{1}(x)+\nu_{2}(x)\big]e^{ik_{a}x} e^{i\theta(x)}+a_{kR}^{\dagger}|g\rangle\langle e|\big[\nu_{1}(x)+\nu_{2}(x)\big]e^{-ik_{a}x} e^{i\theta(x)}\\
&\,\quad+a_{kL}|e\rangle\langle g|\big[\nu_{1}(x)+\nu_{2}(x)\big]e^{-ik_{a}x} e^{-i\theta(x)}+a_{kR}|e\rangle\langle g|\big[\nu_{1}(x)+\nu_{2}(x)\big]e^{ik_{a}x} e^{-i\theta(x)}\Big],
\label{D9}
\end{split}
\tag{S16}
\end{equation}
where $\Lambda$ is the characteristic width and $\int dx \nu_{1,2}(x)=\xi$ is satisfied.
The coupling phase difference is no longer a constant which should be expressed as a function of position $\theta(x)=x\theta_{1}/d$ assuming that the driving field is incident with an unchanged angle.
Then the dynamical equations become
\begin{equation}
\begin{split}
\partial_{t}\tilde{u}^{A_{c}}_{d}(t)&=-(i\omega_{e}-i\Delta_{c1}-i\frac{2\Omega_{c1}^{2}}{\Delta_{c1}}+2\gamma)\tilde{u}^{A_{c}}_{d}(t)-i\int dx\int dk_{a}\big[\nu_{1}(x)+\nu_{2}(x)\big]e^{ik_{a}x}e^{-i\theta(x)}z_{1}(t)\\
&\,\quad-i\int dx\int dk_{a}\big[\nu_{1}(x)+\nu_{2}(x)\big]e^{-ik_{a}x}e^{-i\theta(x)}z_{2}(t),\\
\partial_{t}z_{1}(t)&=-i\omega_{ka}z_{1}(t)-i\tilde{u}^{A_{c}}_{d}(t)\int dx\big[\nu_{1}(x)+\nu_{2}(x)\big]e^{-ik_{a}x}e^{i\theta(x)},\\
\partial_{t}z_{2}(t)&=-i\omega_{ka}z_{2}(t)-i\tilde{u}^{A_{c}}_{d}(t)\int dx\big[\nu_{1}(x)+\nu_{2}(x)\big]e^{ik_{a}x}e^{i\theta(x)}.\\
\label{D10}
\end{split}
\tag{S17}
\end{equation}
With the same procedure, we have
 \begin{equation}
\begin{split}
\partial_{t}\tilde{u}^{A_{c}}_{d}(t)&=-(i\omega_{e}-i\Delta_{c1}-i\frac{2\Omega_{c1}^{2}}{\Delta_{c1}}+2\gamma)\tilde{u}^{A_{c}}_{d}(t)-\frac{i}{v_{g}}\int dx\int d\omega_{ka}\big[\nu_{1}(x)+\nu_{2}(x)\big]e^{i\omega_{ka}x/v_{g}} e^{-i\theta(x)}e^{-i\omega_{ka}t}z_{1}(0)\\
&\,\quad-\frac{i}{v_{g}}\int dx\int d\omega_{ka}[\nu_{1}(x)+\nu_{2}(x)\big]e^{-i\omega_{ka}x/v_{g}} e^{-i\theta(x)}e^{-i\omega_{ka}t}z_{2}(0)\\
&\,\quad-\frac{1}{2v_{g}}\int_{0}^{t}dt'\tilde{u}^{A_{c}}_{d}(t')\int dx \int dx'\int d\omega_{ka}\Big[\nu_{1}(x)\nu_{1}(x')+\nu_{1}(x)\nu_{2}(x')+\nu_{1}(x')\nu_{2}(x)+\nu_{2}(x)\nu_{2}(x')\Big]\\
&\,\quad\times\Big[e^{-i\omega_{ka}[t-t'-|x-x'|/v_{g}]}+e^{-i\omega_{ka}[t-t'+|x-x'|/v_{g}]}\Big] (e^{-i\theta|x-x'|}+e^{i\theta|x-x'|})\\
&=-(i\omega_{e}-i\Delta_{c1}-i\frac{2\Omega_{c1}^{2}}{\Delta_{c1}}+2\gamma)\tilde{u}^{A_{c}}_{d}(t)-\frac{i}{v_{g}}\int dx\int d\omega_{ka}\big[\nu_{1}(x)+\nu_{2}(x)\big]e^{i\omega_{ka}x/v_{g}} e^{-i\theta(x)}e^{-i\omega_{ka}t}z_{1}(0)\\
&\,\quad-\frac{i}{v_{g}}\int dx\int d\omega_{ka}[\nu_{1}(x)+\nu_{2}(x)\big]e^{-i\omega_{ka}x/v_{g}} e^{-i\theta(x)}e^{-i\omega_{ka}t}z_{2}(0)\\
&\,\quad-\frac{\pi}{v_{g}}\int_{0}^{t}dt'\tilde{u}^{A_{c}}_{d}(t')\int dx \int dx'\int d\omega_{ka}\Big[\nu_{1}(x)\nu_{1}(x')+\nu_{1}(x)\nu_{2}(x')+\nu_{1}(x')\nu_{2}(x)+\nu_{2}(x)\nu_{2}(x')\Big]\\
&\,\quad\times\Big[\delta(t'-t+|x-x'|/v_{g})+\delta(t'-t-|x-x'|/v_{g})\Big] (e^{-i\theta|x-x'|}+e^{i\theta|x-x'|})\\
&=-(i\omega_{e}-i\Delta_{c1}-i\frac{2\Omega_{c1}^{2}}{\Delta_{c1}}+2\gamma)\tilde{u}^{A_{c}}_{d}(t)-\frac{i}{v_{g}}\int dx\int d\omega_{ka}\big[\nu_{1}(x)+\nu_{2}(x)\big]e^{i\omega_{ka}x/v_{g}} e^{-i\theta(x)}e^{-i\omega_{ka}t}z_{1}(0)\\
&\,\quad-\frac{i}{v_{g}}\int dx\int d\omega_{ka}[\nu_{1}(x)+\nu_{2}(x)\big]e^{-i\omega_{ka}x/v_{g}} e^{-i\theta(x)}e^{-i\omega_{ka}t}z_{2}(0)\\
&\,\quad-\frac{\pi}{v_{g}}\int_{0}^{t}dt'\tilde{u}^{A_{c}}_{d}(t')\int dx \int dx'\int d\omega_{ka}\Big[\nu_{1}(x)\nu_{1}(x')+\nu_{1}(x)\nu_{2}(x')+\nu_{1}(x')\nu_{2}(x)+\nu_{2}(x)\nu_{2}(x')\Big]\\
&\,\quad\times\Big[\Theta(t-|x-x'|/v_{g})\tilde{u}^{A_{c}}_{d}(t-|x-x'|/v_{g})\Big] (e^{-i\theta|x-x'|}+e^{i\theta|x-x'|}).\\
\label{D11}
\end{split}
\tag{S18}
\end{equation}
After Fourier-transforming this equation as
\begin{equation}
\begin{split}
\tilde{u}^{A_{c}}_{d}(\omega_{ka})&=\frac{\frac{2\pi}{v_{g}}\int dx\big[\nu_{1}(x)+\nu_{2}(x)\big]e^{i\omega_{ka}x/v_{g}} e^{-i\theta(x)}z_{1}(0)+\frac{2\pi}{v_{g}}\int dx\big[\nu_{1}(x)+\nu_{2}(x)\big]e^{-i\omega_{ka}x/v_{g}} e^{-i\theta(x)}z_{2}(0)}{\omega_{ka}-\omega_{e}+\Delta_{c1}+\frac{2\Omega_{c1}^{2}}{\Delta_{c1}}+2i\gamma+F}
\label{D12}
\end{split}
\tag{S19}
\end{equation}
with $F=i\frac{2\pi}{v_{g}}\int dx\int dx'\big[\nu_{1}(x)\nu_{1}(x')+\nu_{1}(x)\nu_{2}(x')+\nu_{1}(x')\nu_{2}(x)+\nu_{2}(x)\nu_{2}(x')\big]e^{i\omega_{ka}|x-x'|/v_{g}} (e^{-i\theta|x-x'|}+e^{i\theta|x-x'|})$,
we replace $\int dx\nu_{1,2}(x)$ to $\sqrt{\frac{v_{g}}{2\pi}}\int d\varphi\tilde{\nu}_{1,2}(\varphi)$ with $\tilde{\nu}_{1}(\varphi)=\frac{\sqrt{\Upsilon_{a}}}{\Lambda}e^{-\frac{2}{\Lambda}|\varphi|}$, $\tilde{\nu}_{2}(\varphi)=\frac{\sqrt{\Upsilon_{a}}}{\Lambda}e^{-\frac{2}{\Upsilon_{1}}|\varphi-(\omega_{ka}/v_{g} \pm\theta_{1}/d)d|}$, and $\varphi=(\omega_{ka}/v_{g} \pm\theta_{1}/d)x$. We have
\begin{equation}
\begin{split}
\tilde{u}^{A_{c}}_{d}(t)&=\int d\omega_{ka}\frac{\frac{1}{\sqrt{2\pi v_{g}}}\left\{\int d\varphi\big[\tilde{\nu}_{1}(\varphi)+\tilde{\nu}_{2}(\varphi)\big]e^{i\omega_{ka}x/v_{g}} e^{-i\theta(x)}z_{1}(0)+\int d\varphi\big[\tilde{\nu}_{1}(\varphi)+\tilde{\nu}_{2}(\varphi)\big]e^{-i\omega_{ka}x/v_{g}} e^{-i\theta(x)}z_{2}(0)\right\}}{\omega_{ka}-\omega_{e}+\Delta_{c1}+\frac{2\Omega_{c1}^{2}}{\Delta_{c1}}+2i\gamma+i(2\Gamma+2iJ+\Gamma_{ex}+ \Gamma'_{ex}+iJ_{ex}+ iJ'_{ex})}
e^{-i\omega_{ka}t},
\label{D13}
\end{split}
\tag{S20}
\end{equation}
where

\begin{equation}
\begin{split}
\Gamma&=\int_{-\infty}^{\infty}d\varphi\int_{-\infty}^{\infty}d\varphi'\tilde{\nu}_{1}(\varphi)\tilde{\nu}_{1}(\varphi')\text{cos}(\varphi-\varphi')=\int_{-\infty}^{\infty}d\varphi\int_{-\infty}^{\infty}d\varphi'\tilde{\nu}_{2}(\varphi)\tilde{\nu}_{2}(\varphi')\text{cos}(\varphi-\varphi')\\
&=\frac{16\Upsilon_{a}}{(\Lambda^{2}+4)^{2}},
\label{D14}
\end{split}
\tag{S21}
\end{equation}

\begin{equation}
\begin{split}
J&=\int_{-\infty}^{\infty}d\varphi\int_{-\infty}^{\infty}d\varphi'\tilde{\nu}_{1}(\varphi)\tilde{\nu}_{1}(\varphi')\text{sin}|\varphi-\varphi'|=\int_{-\infty}^{\infty}d\varphi\int_{-\infty}^{\infty}d\varphi'\tilde{\nu}_{2}(\varphi)\tilde{\nu}_{2}(\varphi')\text{sin}|\varphi-\varphi'|\\&=\frac{\Upsilon_{a}\Lambda(\Lambda^{2}+12)}{(\Lambda^{2}+4)^{2}},
\label{D15}
\end{split}
\tag{S22}
\end{equation}

\begin{equation}
\begin{split}
\Gamma_{ex}&=\int_{-\infty}^{\infty}d\varphi\int_{-\infty}^{\infty}d\varphi'\tilde{\nu}_{1}(\varphi)\tilde{\nu}_{2}(\varphi')\text{cos}(\varphi-\varphi')=\frac{16\Upsilon_{a}\text{cos}[(\omega_{ka}/v_{g} +\theta_{1}/d)d]}{(\Lambda^{2}+4)^{2}},\quad x>x',
\label{D16}
\end{split}
\tag{S23}
\end{equation}

\begin{equation}
\begin{split}
\Gamma'_{ex}&=\int_{-\infty}^{\infty}d\varphi\int_{-\infty}^{\infty}d\varphi'\tilde{\nu}_{1}(\varphi)\tilde{\nu}_{2}(\varphi')\text{cos}(\varphi-\varphi')=\frac{16\Upsilon_{a}\text{cos}[(\omega_{ka}/v_{g} -\theta_{1}/d)d]}{(\Lambda^{2}+4)^{2}},\quad  x<x',
\label{D17}
\end{split}
\tag{S24}
\end{equation}

\begin{equation}
\begin{split}
J_{ex}&=\int_{-\infty}^{\infty}d\varphi\int_{-\infty}^{\infty}d\varphi'\tilde{\nu}_{1}(\varphi)\tilde{\nu}_{2}(\varphi')\text{sin}|\varphi-\varphi'|=\frac{\Upsilon_{a}}{(\Lambda^{2}+4)^{2}}\big\{\big[(8(\omega_{ka}/v_{g} +\theta_{1}/d)d+e^{-\frac{2(\omega_{ka}/v_{g} +\theta_{1}/d)d}{\Lambda}}\\
&\,\quad\times 2\Lambda^{2}(\omega_{ka}/v_{g} +\theta_{0}/d)d+12\Lambda+\Lambda^{3}\big]+16\text{sin}[(\omega_{ka}/v_{g} +\theta_{1}/d)d]\big\},\quad x>x',
\label{D18}
\end{split}
\tag{S25}
\end{equation}

\begin{equation}
\begin{split}
J'_{ex}&=\int_{-\infty}^{\infty}d\varphi\int_{-\infty}^{\infty}d\varphi'\tilde{\nu}_{1}(\varphi)\tilde{\nu}_{2}(\varphi')\text{sin}|\varphi-\varphi'|=\frac{\Upsilon_{a}}{(\Lambda^{2}+4)^{2}}\big\{\big[(8(\omega_{ka}/v_{g}-\theta_{1}/d)d+e^{-\frac{2(\omega_{ka}/v_{g}-\theta_{1}/d)d}{\Lambda}}\\
&\,\quad\times 2\Lambda^{2}(\omega_{ka}/v_{g} -\theta_{1}/d)d+12\Lambda+\Lambda^{3}\big]+16\text{sin}[(\omega_{ka}/v_{g} -\theta_{1}/d)d]\big\},\quad  x<x'.
\label{D19}
\end{split}
\tag{S26}
\end{equation}
For the left-incident photon with $z_{1}(0)\neq0$ and $z_{2}(0)=0$, we substitute this equation into $z_{1}(t)$ as
\begin{equation}
\begin{split}
z_{1}(t)&=e^{-i\omega_{ka}t}z_{1}(0)-i\int d\varphi\int d\varphi'\big[\tilde{\nu}_{1}(\varphi)+\tilde{\nu}_{2}(\varphi)\big]e^{-i\omega_{ka}x/v_{g}} e^{i\theta(x)}\\
&\,\quad\times\int d\omega'_{ka}\frac{\big[\tilde{\nu}_{1}(\varphi')+\tilde{\nu}_{2}(\varphi')\big]e^{i\omega'_{ka}x'/v_{g}} e^{-i\theta(x')}z_{1}(0)}{\omega'_{ka}-\omega_{e}+\Delta_{c1}+\frac{2\Omega_{c1}^{2}}{\Delta_{c1}}+2i\gamma+i(2\Gamma+2iJ+\Gamma_{ex}+ \Gamma'_{ex}+iJ_{ex}+ iJ'_{ex})}e^{-i\omega_{ka}t}
\int_{0}^{t}dt'\tilde{u}^{A_{c}}_{d}(t')e^{i(\omega_{ka}-\omega'_{ka})t'},\\
z_{1}(\infty)&=e^{-i\omega_{ka}t}z_{1}(0)\left\{1-i\frac{\int d\varphi\int d\varphi'[\tilde{\nu}_{1}(\varphi)+\tilde{\nu}_{2}(\varphi)\big][\tilde{\nu}_{1}(\varphi')+\tilde{\nu}_{2}(\varphi')\big]\text{cos}(\varphi-\varphi')}{\omega_{ka}-\omega_{e}+\Delta_{c1}+\frac{2\Omega_{c1}^{2}}{\Delta_{c1}}+2i\gamma+i(2\Gamma+2iJ+\Gamma_{ex}+ \Gamma'_{ex}+iJ_{ex}+ iJ'_{ex})}\right\}\\
&=e^{-i\omega_{ka}t}z_{1}(0)\left[1-\frac{2i(\Gamma+ \Gamma'_{ex})}{\omega_{ka}-\omega_{e}+\Delta_{c1}+\frac{2\Omega_{c1}^{2}}{\Delta_{c1}}+2i\gamma+i(2\Gamma+2iJ+\Gamma_{ex}+ \Gamma'_{ex}+iJ_{ex}+ iJ'_{ex})}\right].
\label{D20}
\end{split}
\tag{S27}
\end{equation}
Therefore, the transmissivity $T^{\rm{eff_{c}}}_{1\rightarrow2}\equiv\frac{|z_{1}(\infty)|^{2}}{|z_{1}(0)|^{2}}$ can be expressed as
\begin{equation}
\begin{split}
T^{\rm{eff_{c}}}_{1\rightarrow2}&=\left|\frac{\delta_{ka}+\Delta_{c1}+\frac{2\Omega_{c1}^{2}}{\Delta_{c1}}+2i\gamma-(2J+J_{ex}+ J'_{ex}) +i\Gamma_{ex}-i\Gamma'_{ex}}{\delta_{ka}+\Delta_{c1}+\frac{2\Omega_{c1}^{2}}{\Delta_{c1}}+2i\gamma+i(2\Gamma+2iJ+\Gamma_{ex}+ \Gamma'_{ex}+iJ_{ex}+ iJ'_{ex})}\right|^{2}.
\label{D21}
\end{split}
\tag{S28}
\end{equation}
With the same procedure, For the right-incident photon with $z_{1}(0)=0$ and $z_{2}(0)\neq0$, we can get the  transmissivity $T^{\rm{eff_{c}}}_{2\rightarrow1}\equiv\frac{|z_{2}(\infty)|^{2}}{|z_{2}(0)|^{2}}$ as
\begin{equation}
\begin{split}
T^{\rm{eff_{c}}}_{2\rightarrow1}&=\left|\frac{\delta_{ka}+\Delta_{c1}+\frac{2\Omega_{c1}^{2}}{\Delta_{c1}}+2i\gamma-(2J+J_{ex}+ J'_{ex}) +i\Gamma'_{ex}-i\Gamma_{ex}}{\delta_{ka}+\Delta_{c1}+\frac{2\Omega_{c1}^{2}}{\Delta_{c1}}+2i\gamma+i(2\Gamma+2iJ+\Gamma_{ex}+ \Gamma'_{ex}+iJ_{ex}+ iJ'_{ex})}\right|^{2}.
\label{D22}
\end{split}
\tag{S29}
\end{equation}

\begin{figure}[pth]
\centering
\includegraphics[width=11 cm]{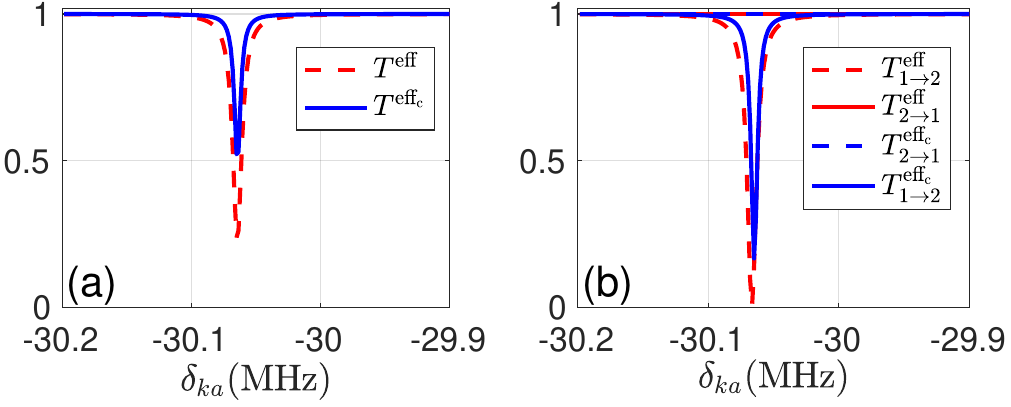}
\caption{Transmissivities $T^{\rm{eff}}_{1\rightarrow2}$ and $T^{\rm{eff}}_{2\rightarrow1}$ with point-like couplings and $T^{\rm{eff_{c}}}_{1\rightarrow2}$ and $T^{\rm{eff_{c}}}_{2\rightarrow1}$ with continuous couplings for (a) $\theta_{1}=\theta_{2}=0$ and (b) $\theta_{1}=\theta_{2}=\pi/2$ with $\phi_{a}=\phi_{b}=\pi/2$ and $\Lambda=\pi/2$.  Other parameters: $\Gamma_{a}=1\,$MHz, $V_{6}=20\,$GHz, $\Omega_{c1}=1\,$MHz, $\Delta_{c1}=30\,$MHz, and $\gamma=1\,$kHz.} 
\label{Appenfig1}
\end{figure}

It is clear that continuous couplings will only change the Lamb shift and effective decay rates, which can be seen from Eqs.~(\ref{D14})-(\ref{D19}).
Figure~\ref{Appenfig1} shows that even considering the continuous coupling, the main results only undergo quantitative changes.
Figures~\figpanel{Appenfig1}{a} and~\figpanel{Appenfig1}{b} correspond to the non-chiral and chiral case respectively indicating that the transmission window and nonreciprocal transmissions still hold.

\section*{II.\quad symmetric frequency conversion with two different fields}

Similarly, for the model shown in Fig. 4 in the main text, we move to the interaction picture of Hamiltonian in Eq.~(10) in the main text as
\begin{equation}
\begin{split}
\mathcal{H}_{B_{k}}(t)&=\int dk_{a}g_{a}a_{k}e^{-i\delta_{ka}t}|r_{1}g_{2} \rangle\langle g_{1}g_{2}|+\int dk_{b}g_{b}b_{k}e^{-i\delta_{kb}t}e^{ik_{b}d}|g_{1}r_{2} \rangle\langle g_{1}g_{2}|+\Omega_{c1}
e^{-i\Delta_{c1}t}|r_{1}r_{2} \rangle\langle r_{1}g_{2}|\\
&\,\quad+\Omega_{c2}e^{-i\Delta_{c2}t}e^{i\theta_{2}}|r_{1}r_{2} \rangle\langle g_{1}r_{2}|+\text{H.c.}.
\end{split}
\tag{S30}
\label{B1}
\end{equation}
So the effective Hamiltonian in the interaction picture can be obtained as
\begin{equation}
\begin{split}
\mathcal{H}_{B_{k}}^{\rm{eff}}(t)
&\simeq\frac{g_{a}^{2}}{\delta_{ka}}\int dk_{a}a_{k}a_{k}^{\dagger}|g_{1}g_{2}\rangle \langle g_{1}g_{2}|+\frac{g_{b}^{2}}{\delta_{kb}}\int dk_{b}b_{k}b_{k}^{\dagger}|g_{1}g_{2}\rangle \langle g_{1}g_{2}|-\left(\frac{\Omega_{c1}^{2}}{\Delta_{c1}}+\frac{\Omega_{c2}^{2}}{\Delta_{c2}}\right)|r_{1}r_{2}\rangle \langle r_{1}r_{2}|-2i\gamma|r_{1}r_{2} \rangle\langle r_{1}r_{2}|\\
&\quad\,+\frac{g_{a}\Omega_{c1}}{\delta_{ka}}\int dk_{a}a_{k}e^{-i(\delta_{ka}+\Delta_{c1})t}|r_{1}r_{2}\rangle
\langle g_{1}g_{2}|+\frac{g_{b}\Omega_{c2}e^{i\theta_{2}}}{\delta_{kb}}\int dk_{b}b_{k}e^{-i(\delta_{kb}+\Delta_{c2})t}e^{ik_{b}d}|r_{1}r_{2}\rangle
\langle g_{1}g_{2}|+\text{H.c.}+\cdots
\end{split}
\tag{S31}
\label{B2}
\end{equation}
and in the Schr\"{o}dinger picture as 
\begin{equation}
\begin{split}
H_{B_{k}}^{\text{eff}}&=(2\omega_{e}+V_{6}-\frac{\Omega_{c1}^{2}}{\Delta_{c1}}-\frac{\Omega_{c2}^{2}}{\Delta_{c2}}-2i\gamma)|r_{1}r_{2} \rangle\langle r_{1}r_{2}|+\int dk_{a}(\omega_{ka}+\omega_{c1})a_{k}^{\dagger}a_{k}+\int dk_{b}(\omega_{kb}+\omega_{c2})b_{k}^{\dagger}b_{k}\\
&\quad\,+\int dk_{a}\xi_{a}a_{k}|r_{1}r_{2}\rangle
\langle g_{1}g_{2}|+\int dk_{b}\xi_{b}b_{k}e^{ik_{b}d}e^{i\theta_{2}}|r_{1}r_{2}\rangle
\langle g_{1}g_{2}|+\text{H.c.}.
\label{B3}
\end{split}
\tag{S32}
\end{equation}
Then the effective Hamiltonian in the real space can be obtained as Eq.~(11) in the main text.

Similarly, by solving the eigenequation $H_{B_{x}}^{\text{eff}}|\tilde{\Psi}_{B_{x}}\rangle=(\omega_{ka}+\omega_{c1})|\tilde{\Psi}_{B_{x}}\rangle$ from Eq.~(11) and Eq.~(12) in the main text, one can obtain
\begin{equation}
\begin{split}
(\omega_{ka}+\omega_{c1})\tilde{\Phi}^{B}_{aR}(x)&=e^{ik_{0}x}\Big(\omega_{0}-iv_{g}\frac{\partial}{\partial x}\Big)\tilde{\Phi}^{B}_{aR}(x)e^{-ik_{0}x}+\xi_{a}\delta(x)\tilde{u}^{B}_{d},\\
(\omega_{ka}+\omega_{c1})\tilde{\Phi}^{B}_{aL}(x)&=e^{-ik_{0}x}\Big(\omega_{0}+iv_{g}\frac{\partial}{\partial x}\Big)\tilde{\Phi}^{B}_{aL}(x)e^{ik_{0}x}+\xi_{a}\delta(x)\tilde{u}^{B}_{d},\\
(\omega_{ka}+\omega_{c1})\tilde{\Phi}^{B}_{bR}(x)&=\Big(\omega_{0}+\omega_{c2}-\omega_{c1}-iv_{g}\frac{\partial}{\partial x}\Big)\tilde{\Phi}^{B}_{bR}(x)+\xi_{b}e^{i\theta_{2}}\delta(x-d)\tilde{u}^{B}_{d},\\
(\omega_{ka}+\omega_{c1})\tilde{\Phi}^{B}_{bL}(x)&=\Big(\omega_{0}+\omega_{c2}-\omega_{c1}+iv_{g}\frac{\partial}{\partial x}\Big)\tilde{\Phi}^{B}_{bL}(x)+\xi_{b}e^{i\theta_{2}}\delta(x-d)\tilde{u}^{B}_{d},\\
(\omega_{ka}+\omega_{c1})\tilde{u}^{B}_{d}&=\Big(2\omega_{e}+V_{6}-2i\gamma\Big)\tilde{u}^{B}_{d}+\xi_{a}[\tilde{\Phi}^{B}_{aR}(0)+\tilde{\Phi}^{B}_{aL}(0)]+\xi_{b}e^{-i\theta_{2}}[\tilde{\Phi}^{B}_{bR}(d)+\tilde{\Phi}^{B}_{bL}(d)].
\label{B4}
\end{split}
\tag{S33}
\end{equation}
Then we substitute the ansatz below (assuming the photon is incident from port 1)
\begin{equation}
\begin{split}
\tilde{\Phi}^{B}_{aR}&=e^{ik_{a}x}[\Theta(-x)+s_{1\rightarrow2}^{\rm{eff}}\Theta(x)],\\
\tilde{\Phi}^{B}_{aL}&=e^{-ik_{a}x}[s_{1\rightarrow1}^{\rm{eff}}\Theta(-x)],\\
\tilde{\Phi}^{B}_{bR}&=e^{ik_{b}x}[s_{1\rightarrow4}^{\rm{eff}}\Theta(x-d)],\\
\tilde{\Phi}^{B}_{bL}&=e^{-ik_{b}x}[s_{1\rightarrow3}^{\rm{eff}}\Theta(-x+d)]
\label{B5}
\end{split}
\tag{S34}
\end{equation}
and can calculate the reflectivity, transmissivity, and backward and forward conversion efficiencies as in Eq.~(13) in the main text.

\section*{III.\quad asymmetric frequency conversion with two different fields}

For the model shown in Fig. 6 in the main text, the Hamiltonian in the interaction picture based on Eq.~(14) can be expressed as 
\begin{equation}
\begin{split}
\mathcal{H}_{C_{k}}(t)&=\int dk_{a}g_{a}a_{k}e^{-i\delta_{ka}t}(|r_{1}g_{2} \rangle\langle g_{1}g_{2}|+e^{ik_{a}d}|g_{1}r_{2} \rangle\langle g_{1}g_{2}|)+\int dk_{b}g_{b}b_{k}e^{-i\delta_{kb}t}(|r_{1}g_{2} \rangle\langle g_{1}g_{2}|+e^{ik_{b}d}|g_{1}r_{2} \rangle\langle g_{1}g_{2}|)\\
&\,\quad+\Omega_{c1}
e^{-i\Delta_{c1}t}(|r_{1}r_{2} \rangle\langle r_{1}g_{2}|+e^{i\theta_{1}}|r_{1}r_{2} \rangle\langle g_{1}r_{2}|)+\Omega_{c2}e^{-i\Delta_{c2}t}(|r_{1}r_{2} \rangle\langle r_{1}g_{2}|+e^{i\theta_{2}}|r_{1}r_{2} \rangle\langle g_{1}r_{2}|)+\text{H.c.}.
\end{split}
\tag{S35}
\label{C1}
\end{equation}
With the same procedure, the effective Hamiltonian in the interaction picture can be obtained as
\begin{equation}
\begin{split}
\mathcal{H}_{C_{k}}^{\rm{eff}}(t)
&\simeq\frac{2g_{a}^{2}}{\delta_{ka}}\int dk_{a}a_{k}a_{k}^{\dagger}|g_{1}g_{2}\rangle \langle g_{1}g_{2}|+\frac{2g_{b}^{2}}{\delta_{kb}}\int dk_{b}b_{k}b_{k}^{\dagger}|g_{1}g_{2}\rangle \langle g_{1}g_{2}|+\frac{2g_{a}g_{b}}{\delta_{ka}}\int dk_{a}a_{k}b_{k}^{\dagger}|g_{1}g_{2}\rangle \langle g_{1}g_{2}|e^{i(\delta_{kb}-\delta_{ka})t}\\
&\quad\,+\frac{2g_{a}g_{b}}{\delta_{kb}}\int dk_{b}b_{k}a_{k}^{\dagger}|g_{1}g_{2}\rangle \langle g_{1}g_{2}|e^{i(\delta_{ka}-\delta_{kb})t}-\frac{2\Omega_{c1}^{2}}{\Delta_{c1}}|r_{1}r_{2}\rangle \langle r_{1}r_{2}|-\frac{2\Omega_{c2}^{2}}{\Delta_{c2}}|r_{1}r_{2}\rangle \langle r_{1}r_{2}|\\
&\quad\,-\frac{2\Omega_{c1}\Omega_{c2}}{\Delta_{c1}}|r_{1}r_{2}\rangle \langle r_{1}r_{2}|e^{i(\Delta_{c1}-\Delta_{c2})t}-\frac{2\Omega_{c1}\Omega_{c2}}{\Delta_{c2}}|r_{1}r_{2}\rangle \langle r_{1}r_{2}|e^{i(\Delta_{c2}-\Delta_{c1})t}\\
&\quad\,+\frac{g_{a}\Omega_{c1}e^{i\theta_{1}}}{\delta_{ka}}\int dk_{a}a_{k}e^{-i(\delta_{ka}+\Delta_{c1})t}(1+e^{ik_{a}d})|r_{1}r_{2}\rangle\langle g_{1}g_{2}|+\frac{g_{b}\Omega_{c2}e^{i\theta_{2}}}{\delta_{kb}}\int dk_{b}b_{k}e^{-i(\delta_{kb}+\Delta_{c2})t}(1+e^{ik_{b}d})|r_{1}r_{2}\rangle\langle g_{1}g_{2}|\\
&\quad\,+\frac{g_{b}\Omega_{c1}e^{i\theta_{1}}}{\delta_{kb}}\int dk_{b}b_{k}e^{-i(\delta_{kb}+\Delta_{c1})t}(1+e^{ik_{b}d})|r_{1}r_{2}\rangle\langle g_{1}g_{2}|+\frac{g_{a}\Omega_{c2}e^{i\theta_{2}}}{\delta_{ka}}\int dk_{a}a_{k}e^{-i(\delta_{ka}+\Delta_{c2})t}(1+e^{ik_{a}d})|r_{1}r_{2}\rangle\langle g_{1}g_{2}|\\
&\quad\,+\text{H.c.}+\cdots.
\end{split}
\tag{S36}
\label{C2}
\end{equation}
Besides omitting the terms related to the single-excitation states done as the last two models, the terms with $e^{\pm i(\Delta_{c1}+\delta_{kb})t}$ and $e^{\pm i(\Delta_{c2}+\delta_{ka})t}$ can also be discarded since they are regarded as high-frequency oscillation terms if assuming $|\Delta_{c1}+\delta_{kb}|\gg g_{b}\Omega_{c1}/\delta_{kb}$ and $|\Delta_{c2}+\delta_{ka}|\gg g_{a}\Omega_{c2}/\delta_{ka}$, while taking $\Delta_{c1}+\delta_{ka}\simeq0$ and $\Delta_{c2}+\delta_{kb}\simeq0$. 
Then, in the same way, after transferring the Hamiltonian to the Schr\"{o}dinger picture, one can obtain the real-space effective Hamiltonian as Eq.~(15) in the main text.

Similarly, by solving the eigenequation one can obtain
\begin{equation}
\begin{split}
(\omega_{ka}+\omega_{c1})\tilde{\Phi}_{aR}^{C}(x)&=e^{ik_{0}x}\Big(\omega_{0}-iv_{g}\frac{\partial}{\partial x}\Big)\tilde{\Phi}_{aR}^{C}(x)e^{-ik_{0}x}+[\xi_{a}\delta(x)+\xi_{a}e^{i\theta_{1}}\delta(x-d)]\tilde{u}_{d}^{C},\\
(\omega_{ka}+\omega_{c1})\tilde{\Phi}_{aL}^{C}(x)&=e^{-ik_{0}x}\Big(\omega_{0}+iv_{g}\frac{\partial}{\partial x}\Big)\tilde{\Phi}_{aL}^{C}(x)e^{ik_{0}x}+[\xi_{a}\delta(x)+\xi_{a}e^{i\theta_{1}}\delta(x-d)]\tilde{u}_{d}^{C},\\
(\omega_{ka}+\omega_{c1})\tilde{\Phi}_{bR}^{C}(x)&=e^{ik_{0}x}\Big(\omega_{0}+\omega_{c2}-\omega_{c1}-iv_{g}\frac{\partial}{\partial x}\Big)\tilde{\Phi}_{bR}^{C}(x)e^{-ik_{0}x}+[\xi_{b}\delta(x)+\xi_{b}e^{i\theta_{2}}\delta(x-d)]\tilde{u}_{d}^{C},\\
(\omega_{ka}+\omega_{c1})\tilde{\Phi}_{bL}^{C}(x)&=e^{-ik_{0}x}\Big(\omega_{0}+\omega_{c2}-\omega_{c1}+iv_{g}\frac{\partial}{\partial x}\Big)\tilde{\Phi}_{bL}^{C}(x)e^{ik_{0}x}+[\xi_{b}\delta(x)+\xi_{b}e^{i\theta_{2}}\delta(x-d)]\tilde{u}_{d}^{C},\\
(\omega_{ka}+\omega_{c1})\tilde{u}_{d}^{C}&=\Big(2\omega_{e}+V_{6}-2i\gamma\Big)\tilde{u}_{d}^{C}+\xi_{a}[\tilde{\Phi}_{aR}^{C}(0)+\tilde{\Phi}_{aL}^{C}(0)]+\xi_{a}e^{-i\theta_{1}}[\tilde{\Phi}_{aR}^{C}(d)+\tilde{\Phi}_{aL}^{C}(d)]\\
&\,\quad+\xi_{b}[\tilde{\Phi}_{bR}^{C}(0)+\tilde{\Phi}_{bL}^{C}(0)]+\xi_{b}e^{-i\theta_{2}}[\tilde{\Phi}_{bR}^{C}(d)+\tilde{\Phi}_{bL}^{C}(d)].
\label{C3}
\end{split}
\tag{S37}
\end{equation}
We substitute the new ansatz below with two coupling points of each waveguide mode (assuming the photon is incident from port 1)
\begin{equation}
\begin{split}
\tilde{\Phi}_{aR}^{C}&=e^{ik_{a}x}\{\Theta(-x)+C_{1}^{\rm{eff}}[\Theta(x)-\Theta(x-d)]+p_{1\rightarrow2}\Theta(x-d)\},\\
\tilde{\Phi}_{aL}^{C}&=e^{-ik_{a}x}\{p_{1\rightarrow1}\Theta(-x)+C_{2}^{\rm{eff}}[\Theta(x)-\Theta(x-d)]\},\\
\tilde{\Phi}_{bR}^{C}&=e^{ik_{b}x}\{C_{3}^{\rm{eff}}[\Theta(x)-\Theta(x-d)]+p_{1\rightarrow4}\Theta(x-d)\},\\
\tilde{\Phi}_{bL}^{C}&=e^{-ik_{b}x}\{p_{1\rightarrow3}\Theta(-x)+C_{4}^{\rm{eff}}[\Theta(x)-\Theta(x-d)]\}.
\label{C4}
\end{split}
\tag{S38}
\end{equation}
The reflectivity, transmissivity, and backward and forward conversion efficiencies can be calculated as in Eq.~(16) in the main text.

\end{widetext}

\end{document}